%% file: main.tex
\documentclass[superscriptaddress, twocolumn, prx, longbibliography]{revtex4-2}
\usepackage[T1]{fontenc}
\usepackage{libertine}
\usepackage{color, amsfonts, amsmath, graphicx, tikz, amssymb, mathtools}
\usepackage{amsthm}
\usepackage[libertine, cmintegrals, bigdelims, vvarbb]{newtxmath}
\usepackage{appendix}
\usepackage{comment}
\usepackage{upgreek}
\definecolor{linkcolor}{rgb}{0,0,0.6}

\definecolor{lgreen} {RGB}{180,210,100}
\definecolor{dblue}  {RGB}{20,66,129}
\definecolor{jblue}  {RGB}{20,50,100}
\definecolor{nblue}  {RGB}{0,120,200}
\definecolor{dgreen} {RGB}{78,138,21}
\definecolor{ngreen} {RGB}{98,158,31}
\definecolor{lred}   {RGB}{220,0,0}
\definecolor{nred}   {RGB}{224,0,0}
\usepackage[pdftex,colorlinks=true,
	pdfstartview = FitV,
	linkcolor    = red,
	citecolor    = blue,
	urlcolor     = blue,	
	hyperindex   = true,
	hyperfigures = false]{hyperref}
\usepackage{cleveref}

\input{abbrev}

\patchcmd{\subsubsection}{\itshape}{\bfseries}{}{}

\usepackage{orcidlink}
\usepackage{dcolumn}
\usepackage{bm}
\usepackage{algorithm}
\usepackage{algpseudocode}

\begin{document}
\title{Thermodynamic Length in Stochastic Thermodynamics of Far-From-Equilibrium Systems: Unification of Fluctuation Relation and Thermodynamic Uncertainty Relation}
\author{Atul Tanaji Mohite\,\orcidlink{0009-0004-0059-1127}}
\email{atul.mohite@uni-saarland.de}
\affiliation{Department of Theoretical Physics and Center for Biophysics, Saarland University, Saarbrücken, Germany}

\author{Heiko Rieger\,\orcidlink{0000-0003-0205-3678}}
\affiliation{Department of Theoretical Physics and Center for Biophysics, Saarland University, Saarbrücken, Germany}

\begin{abstract}
The Boltzmann distribution for an equilibrium system constrains the statistics of the system by the energetics. Despite the non-equilibrium generalization of the Boltzmann distribution being studied extensively, a unified framework valid for far-from-equilibrium discrete state systems is lacking. Here, we derive an exact path-integral representation for discrete state processes and represent it using the exponential of the action for stochastic transition dynamics. Solving the variational problem, the effective action is shown to be equal to the inferred entropy production rate (a thermodynamic quantity) and a non-quadratic dissipation function of the thermodynamic length (TL) defined for microscopic stochastic currents (a dynamic quantity). This formulates a far-from-equilibrium analog of the Boltzmann distribution, namely, the minimum action principle. The non-quadratic dissipation function is physically attributed to incorporating non-Gaussian fluctuations or far-from-equilibrium non-conservative driving. Further, an exact large deviation dynamical rate functional is derived. The equivalence of the variational formulation with the information geometric formulation is proved. The non-quadratic TL recovers the non-quadratic thermodynamic-kinetic uncertainty relation (TKUR) and the speed limits, which are tighter than the close-to-equilibrium quadratic formulations. Moreover, if the transition affinities are known, the non-quadratic TL recovers the fluctuation relation (FR). The minimum action principle manifests the non-quadratic TKUR and FR as two faces corresponding to the thermodynamic inference and partial control descriptions, respectively. In addition, the validity of these results is extended to coarse-grained observable currents, strengthening the experimental/numerical applicability of them. 
\end{abstract}

\date{\today}

\maketitle
\section{Introduction}\label{sec:introduction}
The Boltzmann distribution is the most fundamental principle in Statistical Physics. It formulates an equivalence between thermodynamics and statistics for equilibrium systems, valid in the thermodynamic limit \cite{boltzmann_1964}. Finite-size/particle systems prone to non-equilibrium fluctuations are ubiquitous and violate the assumption of the thermodynamic limit. By relaxing the assumption of the thermodynamic limit, the framework of stochastic thermodynamics (ST) enables to define thermodynamic quantities for the stochastic transition of a microscopic system \cite{seifert_2012, sekimoto, Shiraishi_2023_book}. In ST, the thermodynamic dissipation cost to sustain non-equilibrium fluctuations and/or driving is quantified by the entropy production rate (EPR). Recently, ST has been extended to `non-reciprocal' systems that violate `actio-reactio' symmetry, and to coarse-grained macroscopic systems \cite{ATM_2024_nr_st,ATM_2024_nr_cg} due to an exact thermodynamically-consistent coarse-graining of microscopic systems \cite{ATM_2024_nr_cg}. This has cemented the applicability of ST to experimentally/practically relevant real-world systems, such as non-ideal particles and coarse-grained mesoscopic/macroscopic degrees of freedom. 

The fluctuation relation (FR) is a fundamental seminal law in ST, which connects the time-reversal asymmetry of dynamics to the stochastic thermodynamic cost \cite{seifert_2012, sekimoto, Shiraishi_2023_book,ATM_2024_nr_st, schnakenberg_1976, Bochkov_1977, Bochkov_1979, Evans_1993, Evans_1994, Jarzynski_1997, Jarzynski_1997_pre, Crooks_1999, Tasaki_2000, Crooks_2000, Maes_2003, Gallavotti_1995, Lebowitz_1999, Maes_2003, Kurchan_1998, Sekimoto_1997, Sekimoto_1998, Seifert_2005, sekimoto, Andrieux_2007, Andrieux_2007_single_current_FT, Touchette_2009}. The first-order mean-field approximation of FR recovers the second law of thermodynamics (an approximate law). Recently, the Thermodynamic-Kinetic Uncertainty Relation (TKUR) has revealed a lower bound on thermodynamic dissipation (a thermodynamic quantity) using the current precision (a dynamic quantity) \cite{Gilmore_1985, Uffink_1999, Horowitz_2020, Barato_2015, Gingrich_2016, Horowitz_2017, Terlizzi_2019, Kwon_2023_TUR_unified}. TKUR obtains a tighter lower bound on the thermodynamic dissipation required to sustain a non-equilibrium process than the second law of thermodynamics. TKUR's relation to Speed Limits (SL) has been explored \cite{ Van_vu_2023, Vo_2022, Vo_2020, Lee_2022, Ito_2018}. FR and TKUR are often treated as distinct fundamental laws in ST, and their connection is missing. TKUR has been derived using FR \cite{Merhav_2010,Hasegawa_2019,Timpanaro_2019,Francica_2022}, but the lower bound obtained on dissipation was loose \cite{Merhav_2010,Hasegawa_2019,Timpanaro_2019,Francica_2022}. Although TKUR has a practical advantage for thermodynamic inference, in contrast to FR, the fundamental/seminal origin of TKUR is debatable, which we will prove. 

Non-equilibrium generalizations of the Boltzmann distribution have been explored extensively \cite{Klein_1954, Callen_1957, Glansdorff_1971, Jaynes_1980, Struchtrup_1998, Qian_2002, Evans_2004, Evans_2005, Lecomte_2005, Martyushev_2006, Wang_2006, Bruers_2007, Bruers_2007_maxEP_minEP, Lecomte_2007, Yoshida_2008, Baule_2008,Touchette_2009, Baule_2010, Martyushev_2010, Niven_2010, Kawazura_2010, Doi_2011, Monthus_2011, Kawazura_2012, Chetrite_2013, Presse_2013, Endres_2017}, and its applications to biological systems are studied \cite{Bialek_2000, Sasai_2003, Lan_2006, Vellela_2009}. 
%
%
{However, the non-equilibrium generalization of the Boltzmann distribution has three major drawbacks. First, a Gaussian approximation for fluctuations/driving, which is identified by a quadratic relation between EPR and driving forces/fluctuations \cite{Bodineau_2004,Derrida_2007,Prados_2011,Hurtado_2011,Espigares_2013,Bertini_2015,Qian_2020}. The Gaussian approximation for fluctuations was originally derived to study close-to-equilibrium (cEQ) systems \cite{Onsager_1953,Onsager_1953_2} and extended to path-integral formulism around the mean-field description \cite{Martin_1973,Janssen_1976,Bausch_1976,Dedominicis_1976}: a \textit{top-down approach} towards state-space fluctuations. However, non-Gaussian fluctuations are important for far-from-equilibrium (fEQ) or finitely small size systems. Notable non-Gaussian state-space formulations assume a specific form of the non-Gaussian noise rather than its microscopic derivation; again, a \textit{top-down approach} to state-space fluctuations \cite{Baule_2023,Huang_2025}. Second, except for a few notable exceptions, existing works focus on the state-space representation of fEQ processes. However, stochasticity in physical systems originates from microscopic transitions, which are the most fundamental level of description. Moreover, state-space formulations fail to distinguish between multiple transition channels connecting the same states, for example, those arising from coupling to different thermodynamic reservoirs or from distinct effective driving mechanisms. Therefore, a transition-space representation of fEQ systems is necessary, since, a state-space formulation is a cEQ methodology that does not fully capture the physical aspects of transitions in fEQ systems. Third, a coherent and unified description of fEQ systems grounded in a single underlying principle is missing, due to contradictions between different formulations, which is again attributed to a phenomenological \textit{top-down approach} towards fEQ principles.
}   

In this work, we derive the minimum action principle for the EPR of discrete state processes \cite{atm_2024_var_epr}. To this end, we use the second quantization method, namely, the Doi-Peliti field theory (DPFT), which preserves non-Gaussian transition fluctuations due to its \textit{bottom-up construction} \cite{Doi_1976,Doi_1976_2,Peliti, Peliti_1986,Weber_2017,ATM_2024_nr_cg}. We derive an exact transition probability measure for discrete state processes, which is equal to the exponential of the action. Hence, a variational formulation for discrete state processes is formulated, namely, the `Minimum Action Principle' (MinAP). We prove that the effective action Lagrangian is equivalent to inferred EPR (a thermodynamic quantity). The Lagrangian is shown to be a non-quadratic function of the cumulants of the microscopic stochastic transition currents (a dynamic quantity) and quantifies the thermodynamic length of stochastic currents. The threefold equality between the transition probability measure, inferred EPR, and current cumulants formulates a far-from-equilibrium analogue of the Boltzmann distribution. 

Using the thermodynamic length (TL) \cite{Salamon_1983,Salamon_1985,Schlogl_1985,Crooks_2007,Ito_2018,Yoshimura_2021,Yoshimura_2021_ig_crn,Loutchko_2022_prr_riemanian}, we demonstrate that the variational formulation yields a non-quadratic TKUR, which provides a tighter bound than the quadratic TKUR. If the transition affinities are known (a partial control description), the Lagrangian reduces to the bilinear form of EPR, which recovers FR for microscopic stochastic currents. Using TL, the non-quadratic TKUR and FR are unified within MinAP; they correspond to the thermodynamic inference and partial control descriptions, respectively. Further, we derive the exact large deviation functional for discrete state processes and discuss its importance compared to the Gaussian and Hessian approximations of the large deviation functional. We extend the validity of MinAP for coarse-grained observable stochastic currents. To this end, we solve the variational problem for the Lagrangian under the constraint imposed by the observable currents. We prove that MinAP is extended to coarse-grained observable stochastic currents with the same underlying framework/structure as the one derived for microscopic transition currents. 

Moreover, the variational formulation broadens the numerical applicability of MinAP \cite{Eyink_1996}, where an exact analytical solution is not feasible, which is usually the case beyond a few exactly solvable prototypical models. For example, variational formulations have been utilized to study the non-equilibrium dynamics: phase transitions, first-passage times, metastability in stochastic systems \cite{Weinan_2004,Heymann_2008,Vanden-Eijnden_2008,Bressloff_2015,Grafke_2017,Grafke_2019,Gagrani_2023,Zakine_2023,Smith_2024}, applications in machine learning \cite{Cherukuri_2017,Lin_2020,Yan_2022}, and for thermodynamic inference in ST \cite{kim_2020,Otsubo_2020,Otsubo_2022,Horiguchi_2024,Tottori_2024}. Hence, the variational formulation of far-from-equilibrium discrete state systems allows us to explore its applications in the future and broadens the practical applicability of ST. As an application of MinAP, assuming `full control' of transition affinities and mobilities, we develop \textit{Generalized finite-time optimal control framework} in Ref.\cite{atm_2025_gftoc}.
\section{Minimum Action Principle}\label{sec:min_action_pr}
\subsection{Setup}\label{sec:setup}
\subsubsection{Thermodynamically-consistent discrete state processes and graphs} 
The discrete-state systems are represented by a graph, such that the state and transition form the nodes and directional edges of the graph, respectively \cite{schnakenberg_1976}. The set of all states is denoted by $\{i\}$. The state probability/density and transition between the states are denoted by $\rho_i$ and $\gamma$, respectively. For each forward unidirectional transition $\gamma$ between reactant state $\rho_{r_\gamma}$ to product state $\rho_{p_\gamma}$, there exists a backward unidirectional transition $-\gamma$ \footnote{ We consider systems with all bidirectional transitions with positive transition rates, this is to ensure the absolute continuousness of the forward path probability measure with respect to the conjugate path probability measure. }. 
{
The state probability and density on the graph correspond to the physical descriptions provided by Markov jump processes and linear chemical reaction networks, respectively. The intermediate coarse-graining step between them and the resulting hydrodynamic description is elucidated in Ref.~\cite{ATM_2024_nr_cg}.
}
The transition currents for the forward and backward reactions are denoted by $j_\gamma = \rho_{r_\gamma} k_\gamma$ and $j_{-\gamma} = \rho_{p_\gamma} k_{-\gamma}$, respectively, where $k_\gamma$ and $k_{-\gamma}$ are the forward and backward transition rates. The set of all unidirectional and bidirectional transitions of a graph is denoted by $\{ \gamma^{\rightharpoonup} \}$ and $\{ \gamma^{\rightleftharpoons} \}$, respectively.

The thermodynamic consistency condition for the transition currents implies that they satisfy the Local Detailed Balance condition (LDB), $ A_{\gamma} = \ln{(j_{\gamma}/j_{-\gamma})} = F_\gamma - \Delta_{\gamma} E + \Delta_{\gamma} S^{state} $. Here, $A_\gamma$ is the transition affinity that quantifies the time-reversal asymmetry or the directionality of the transition currents. It is decomposed into three terms: an external non-conservative non-equilibrium driving $F_{\gamma}$ supported by a thermodynamic reservoir, change in the equilibrium energy functional $\Delta_{\gamma}E$ due to transition $\gamma$, and the change in the state entropy $S_{i}^{state}= - \ln{ \left( \rho_i \right) } $ \cite{seifert_2012}. Here, the conservative force is represented as a change of chemical potential ($\mu$) between the reactant and product state $-\Delta_{\gamma}E = \mu_{r_\gamma} - \mu_{p_\gamma}$ \cite{ATM_2024_nr_st}. The energy $E$ is controlled by the set of parameters $\{\lambda\}$. The equilibrium Boltzmann distribution is given by $\rho_i^E = e^{-E_i + \psi_E}$, with the corresponding equilibrium free energy $\psi_E = - \ln{ \left( \sum_{\{i\}} e^{-E_i} \right) }$. 
{
The formulation of the LDB condition in ST relies on timescale separation and weak coupling between the system degrees of freedom and the environment. Nevertheless, LDB allows one to exactly quantify the thermodynamic cost associated with any stochastic change of the system’s state. This is of paramount importance for establishing the physical connection between the statistical and thermodynamic properties of fEQ systems, much like the Boltzmann distribution does for equilibrium systems. However, the derivation of MinAP is valid irrespective of the LDB condition.
}

The total bidirectional current and traffic for the bidirectional transition $\gamma^\rightleftharpoons$ are defined as $J_\gamma = j_\gamma - j_{-\gamma}$ and $T_\gamma = j_\gamma + j_{-\gamma}$. They correspond to the decomposition of bidirectional currents into linearly independent time-antisymmetric and time-symmetric parts, respectively. The scaled traffic physically quantifies the variance of the current, with a large deviation scaling parameter that plays a role similar to the inverse temperature for equilibrium systems \cite{ATM_2024_nr_cg,ATM_2024_nr_st,Maes_2020}. For example, the observation time $\tau$ for dynamical systems \cite{Maes_2017,Maes_2020,Touchette_2009}, and the system volume $\mathcal{V}$ for fluctuating hydrodynamic description \cite{Bertini_2015,Touchette_2009,ATM_2024_nr_cg,ATM_2024_nr_st} are the relevant large deviation scaling parameters. Defining the mobility $D_\gamma = \sqrt{ j_{\gamma} j_{-\gamma} } = \sqrt{ \rho_{p_\gamma} \rho_{r_\gamma}  e^{\mu_{p_\gamma} + \mu_{r_\gamma}}}$ for $\gamma^\rightleftharpoons$, mean currents and traffics satisfy relations with affinities and mobilities, $ J_\gamma = 2 D_\gamma \sinh{(A_\gamma/2)}$ and $ T_\gamma = 2 D_\gamma \cosh{(A_\gamma / 2)}$, which justifies the nomenclature of mobility. 
\subsubsection{Dynamics} 
The continuity equation constrains the temporal state-space flow generated due to transitions; it reads,  
\begin{equation}\label{eq:master_equation}
\begin{split}
    \partial_t \vec{\rho} = \mathbb{S} \vec{J}_{},
\end{split}    
\end{equation}
where $\mathbb{S}$ is the stoichiometry matrix for the graph that dictates the contraction from the transition space to the state space \cite{schnakenberg_1976}. The states and transitions are represented in the column vector notation $\vec{\rho} = (.., \rho_i, ..)^T$ and $\vec{J} = (.., J_\gamma, ..)^T$. The state-space and transition-space have dimensions $|\{i\}|$ and $|\{\gamma^\rightleftharpoons\}|$, respectively. Hence, $\mathbb{S}$ is a matrix of dimension $|\{i\}| \times |\{\gamma^\rightleftharpoons\}|$. Here, $||$ denotes the dimension of the set. For a transition $\gamma^{\rightleftharpoons}$, $\mathbb{S}_\gamma$ is the row vector corresponding to the transition $\gamma^\rightleftharpoons$. The negative and positive entries of $\mathbb{S}_\gamma$ correspond to the reactant and product species vectors, respectively. For a graph, $\vec{r}_\gamma$ and $\vec{p}_\gamma$ have only one non-vanishing entry denoted by index ${r}_\gamma$ and ${p}_\gamma$ for the reactant and product, respectively. 
{The reactant and product indices ${r}_\gamma$ and ${p}_\gamma$ for each transition $\gamma$ dictates the contraction from the transition-space $(\gamma \in \{\gamma^\rightleftharpoons\})$ to the state-space $({r}_\gamma, {p}_\gamma \in \{i\})$. }

{To highlight the importance of the transition-space representation of fEQ systems discussed in the introduction, let us consider two conformations of a protein `1' and `2', with transitions occurring between them via an equilibrium thermodynamic reservoir and an enzymatic thermodynamic reservoir that favors the product formation through an effective non-equilibrium driving $F_{enz}$. For this prototypical system, $\{i\} = \{1, 2\}$ and $\{ \gamma^\rightleftharpoons \} = \{eq, enz \}$. Both transition channels connect the same pair of states, such that $r_{eq} = 1, p_{eq} = 2$ and $r_{enz} = 1, p_{enz} = 2$. But the thermodynamic cost associated with the transition depends on the channel taken: $A_{eq} = \mu_1 - \mu_2 + \ln{(\rho_1)} - \ln{(\rho_2)}$ and $A_{enz} = F_{enz} + \mu_1 - \mu_2 + \ln{(\rho_1)} - \ln{(\rho_2)}$. Therefore, the transition space representation of the microscopic dynamics preserves the physical (both dynamical and thermodynamic) distinction between different transition channels connecting the same states. Hence, throughout this work, we employ the transition space representation. If a state space description is required, one may invoke the continuity equation to contract the dynamics from transition-space to state-space.
}
\subsubsection{Orthogonal decomposition of thermodynamic dissipation for graphs}
The mean EPR $\langle \dot{\Sigma} \rangle$ for the graph has a bilinear form, namely, force times current \cite{schnakenberg_1976},
\begin{equation}\label{eq:mean_EPR}
    \langle \dot{\Sigma} \rangle = \sum_{ \{\gamma^\rightleftharpoons\} } J_\gamma A_\gamma,
\end{equation}
$\langle \dot{\Sigma} \rangle$ is further decomposed into its three linearly independent contributions, 
\begin{equation}\label{eq:mean_EPR}
\begin{split}
    -\dot{\psi}_E & = -  \dot{\lambda} \: \partial_\lambda \psi_E,
    \\
    \langle \dot{\Sigma}_E^{ex} \rangle & = - \sum_{\{i\}} d_t \rho_i \ln{\left( \frac{\rho_i}{\rho_i^E} \right)} = -d_t D_E^{KL},
    \\
    \langle \dot{\Sigma}^{hk} \rangle & = \sum_{ \{ \gamma^\rightleftharpoons \} } T_\gamma^\perp F_\gamma \sinh{ \left( \frac{F_\gamma}{2} \right) },
\end{split}
\end{equation}
the quasistatic driving work rate ($-\dot{\psi}_E$), the excess EPR $(\langle \dot{\Sigma}^{ex} \rangle)$ and the housekeeping EPR $(\langle \dot{\Sigma}^{hk} \rangle)$, respectively. They physically correspond to the instantaneous change in free energy implemented by changing control parameters, the statistical distance of instantaneous state distribution $\{ \rho_i \}$ from the Boltzmann distribution for states $\{ \rho_i^E \}$, and the thermodynamic cost of maintaining non-conservative driving $\{F_\gamma\}$, respectively. They are defined in $\{ \lambda \}$, $\{i\}$ and $\{ \gamma^\rightleftharpoons \}$ space, respectively. $-\dot{\psi}_E$ and $\langle \dot{\Sigma}_E^{ex} \rangle$ do not require information on the transition topology. Thus, they correspond to dissipation terms that are integrated using \cref{eq:master_equation} to obtain boundary terms for EPR in the control parameter space and state-space, respectively. 

The excess-housekeeping decomposition of the EPR is formulated by identifying the conservative and non-conservative decompositions of the transition affinity, which satisfy time-reversal symmetry and anti-symmetry, respectively. The decomposition of the affinity is $A_\gamma = F_\gamma +\Delta_\gamma S_i^{state/E} = A_\gamma^{nc} + A_\gamma^{rel}$ with the relative state entropy defined with respect to the Boltzmann distribution $S_i^{state/E} = - \ln{ \left( \rho_i/ \rho_i^E \right) }$ for fixed values of $\{ \lambda \}$. $A_\gamma^{rel}$ and $A_\gamma^{nc}$ are transition affinities that generate instantaneous (short-time) and steady-state dynamics given by \cref{eq:master_equation}. Physically, this symmetry is a manifestation of the short-time relaxation and long-time steady-state symmetry of currents. This symmetry generates an orthogonal decomposition of EPR, and $\langle \dot{\Sigma}^{hk} \rangle $ is simplified to, $\langle \dot{\Sigma}^{hk} \rangle = \sum_{\{\gamma^\rightleftharpoons\}} F_\gamma J_\gamma^a$, where $J_\gamma^a$ is the anti-symmetric part of $J_\gamma$ under the orthogonal transformation, $F_\gamma \to -F_\gamma$ \cite{ATM_2024_nr_st,ATM_2024_nr_cg}. This resolves to $\langle \dot{\Sigma}^{hk} \rangle = \sum_{ \{\gamma^\rightleftharpoons\} } T_\gamma^{\perp} F_\gamma \sinh{(F_\gamma/2)}$ in \cref{eq:mean_EPR}, where $T_\gamma^\perp = T_\gamma|_{F_\gamma=0}$ is the equilibrium traffic obtained by plugging in $F_\gamma = 0$ (the direction orthogonal to non-conservative driving). This implies that the mean housekeeping EPR is equal to the equilibrium traffic multiplied by a non-quadratic function of the non-conservative driving force $F_\gamma$ \cite{ATM_2024_nr_st,ATM_2024_nr_cg}. Putting in the steady-state distribution as the reference Boltzmann distribution $E = ss$, one obtains the usual adiabatic-non-adiabatic decomposition \cite{seifert_2012}, a sub-case of the orthogonal decomposition.
\subsection{Variational formulation}\label{sec:var_formulation}
\subsubsection{Derivation using DPFT }\label{sec:var_dp_theory}
We derive the exact path-integral representation of discrete-state processes using an exact second-quantized approach that preserves non-Gaussian transition statistics, namely, Doi-Peliti field theory (DPFT) \cite{Doi_1976,Doi_1976_2,Peliti,Peliti_1986,Weber_2017,ATM_2024_nr_cg}.

\textit{Microscopic transitions represented as a second-quantized Hamiltonian}. \textemdash \:
In DPFT, the second-quantized representation for the creation and annihilation of state $i$ is given by creation and annihilation operators $\hat{\eta}^\dagger_{i}$ and $\hat{\eta}_{i}$, respectively. Hence, $\hat{\eta}^{\dagger}_{p_\gamma}$ and $\hat{\eta}_{{r}_\gamma}$ correspond to the creation of the product state and the annihilation of the reactant state for the transition $\gamma^\rightharpoonup$. The corresponding inverse transition $\gamma^\leftharpoondown$ is represented in the second-quantized form by $\hat{\eta}^{\dagger}_{r_\gamma}$ and $\hat{\eta}_{{p}_\gamma}$ for the creation of the reactant state and the annihilation of the product state. Using Doi-Peliti field theory, the second-quantized Hamiltonian operator for the bidirectional transition $\gamma^\rightleftharpoons$ reads \cite{Doi_1976,Doi_1976_2,Peliti,Peliti_1986,Weber_2017,ATM_2024_nr_cg}:
\begin{equation}\label{eq:hamiltonian_quantised_single_reactive_jump}
\begin{split}
\hat{H}_{\gamma} [ \crtlong{\mathbf{r}_\gamma}, \crtlong{\mathbf{p}_\gamma}, \anhlong{\mathbf{r}_\gamma}, \anhlong{\mathbf{p}_\gamma} ]
&= \left( \crtlong{\mathbf{r}_\gamma} - \crtlong{\mathbf{p}_\gamma} \right)\left( \anhlong{\mathbf{r}_\gamma} k_{\gamma} - \anhlong{\mathbf{p}_\gamma} k_{-\gamma} \right).
\end{split}    
\end{equation}
The transition rates $k_\gamma$ and $k_{-\gamma}$ are assumed not to depend on $\rho_i$. A state dependence of the transition rates $\{k_\gamma\}$ would create a technical sophistication in implementing DPFT, discussed later \cite{ATM_2024_nr_cg}. It can be addressed by a careful implementation of DPFT for the state-dependent $\{k_{\gamma}\}$ and represents a technical intricacy rather than a conceptual one \cite{ATM_2024_nr_cg}. The second-quantized Hamiltonian for all linearly independent microscopic transitions reads \cite{Doi_1976,Doi_1976_2,Peliti,Peliti_1986,Weber_2017,ATM_2024_nr_cg}:
\begin{equation}\label{eq:hamiltonian_quantized_crn}
    \hat{H} [ \{ \hat{\eta}_i^\dagger, \hat{\eta}_i \} ] = \sum_{\{ \gamma^\rightleftharpoons \}} \hat{H}_{\gamma} [ \crtlong{\mathbf{r}_\gamma}, \crtlong{\mathbf{p}_\gamma}, \anhlong{\mathbf{r}_\gamma}, \anhlong{\mathbf{p}_\gamma} ].
\end{equation}
\vspace{3pt}

\textit{Coherent state and master equation}. \textemdash \:
The coherent state $ | \phi_{i} \rangle $ and its corresponding dual state $\langle \phi^*_{i} | $ are defined as \cite{Doi_1976,Doi_1976_2,Peliti,Peliti_1986,Weber_2017,ATM_2024_nr_cg}:
\begin{equation}\label{eq:coherant_state_single_species}
\begin{split}
    | \phi_{i} \rangle 
    =  \sum_{l} \frac{ \left( \phi_i \right)^{l} \left( \crtlong{i} \right)^{l}}{ l !}  | 0 \rangle_i, \hspace{0.5cm}
    \langle \phi^*_{i} | 
    = \:_i\langle 0 | \sum_{l} \frac{ \left( \phi_i^* \right)^{l} \left( \anhlong{i} \right)^{l}}{ l!}. 
\end{split}
\end{equation}
Here, $\phi_i$ is the eigenvalue of the coherent state with complex conjugate $\phi_i^*$ and satisfies $\anhlong{i} | \phi_{i} \rangle = \phi_i | \phi_{i} \rangle $ and $\langle \phi_i | \crtlong{i} = \langle \phi_{i} | \phi_i^*$. The physical interpretation of the coherent state \cref{eq:coherant_state_single_species} is visualized using its alternative representation, $| \phi_i \rangle  = \sum_{l} P_i (l) | l_i \rangle$, which gives the sum over a Poissonian probability distribution for the state occupancy $l$ for state $i$ $\rho_i$ with $P_i(l) = (\phi_i)^l/l!$. Then, the master \cref{eq:master_equation} for the state probability flow using the coherent state has the second-quantized form $\partial_t | \{ \phi_i \} \rangle = -\hat{H} [ \{ \hat{\eta}_i^\dagger, \hat{\eta}_i \} ] | \{ \phi_i \} \rangle$ \cite{Doi_1976,Doi_1976_2,Peliti,Peliti_1986,ATM_2024_nr_cg,Weber_2017}.

Therefore, the inner product of an operator $\hat{O}$ using the coherent state is equivalent to computing the expectation value over a Poissonian probability measure for the state occupancy ($\mathcal{O} [\{\phi_i^*, \phi_i\}] = \langle \phi_i^*|\hat{O} [ \{ \hat{\eta}_i^\dagger, \hat{\eta}_i \} ] | \phi_i \rangle / \langle \phi_i^* | \phi_i \rangle$). The coherent state and its dual are eigenvectors of $\anhlong{i} | \phi_{i} \rangle = \phi_i | \phi_{i} \rangle $ and $\langle \phi_i | \crtlong{i} = \langle \phi_{i} | \phi_i^*$. Thus, computing the expectation value of the operator over Poissonian occupancy of the states is equivalent to replacing $\hat{\eta}_i \to \phi_i$ and $\hat{\eta}_i^\dagger \to \phi_i^*$ if $\hat{O} [ \{ \hat{\eta}_i^\dagger, \hat{\eta}_i \} ]$ is normal ordered \cite{ATM_2024_nr_cg}. This highlights the reason for choosing transition rates $\{k_{\gamma}\}$ that are independent of the state occupancies $\{\rho_i\}$, since it adds another sophistication to obtain a normal-ordered expression for $\{k_{\gamma}\}$ in the transition Hamiltonian \cref{eq:hamiltonian_quantised_single_reactive_jump} \cite{ATM_2024_nr_cg}. Using the tensor product, the coherent state $| \{ \phi_{i} \} \rangle = \prod_{} \otimes | \phi_{i} \rangle $ and its conjugate dual state $\langle \{ \phi^*_{i} \} | = \prod_{}\otimes \langle \phi_i | $ for the set of all microstates are defined. 

\vspace{3pt}

\textit{The mesoscopic Doi-Peliti Action, Lagrangian and Hamiltonian in the creation-annihilation picture}. \textemdash \:
Using DPFT \cite{Doi_1976,Doi_1976_2,Peliti,Peliti_1986,ATM_2024_nr_cg,Weber_2017}, the exact `stochastic' path integral formulation for the transition probability measure for the discrete-state process written using the eigenvalues of the coherent state reads \footnote{The equality in \cref{eq:doi_peliti_path_integral_coherant_state_picture} for the transition probability measure is trivially assumed by incorporating the probability normalization factor, $\int \mathbb{ D } \{ \phi_{i}^* \} \: \mathbb{ D } \{ \phi_i \} \mathcal{P} \left[ \left\{ \phi_{i}^*, \phi_{i} \right\} \right] = 1$. Here, $\mathbb{ D } \{ \phi_{i}^* \} $ and $ \mathbb{ D } \{ \phi_i \}$ denote the path integral over all coherent-state eigenvalues. Henceforth, we omit the normalization constant for the transition probability measure; however, its inclusion for probability conservation is presumed.}, 
\begin{equation}\label{eq:doi_peliti_path_integral_coherant_state_picture}
\begin{split}
    \mathcal{P} \left[ \left\{ \phi_{i}^*, \phi_{i} \right\} \right] 
    = \: e^{ -\mathcal{S}_{DP} \left[ \left\{ \phi_{i}^*, \phi_{i} \right\} \right] },
\end{split}
\end{equation}
Using $\phi_i$, $\phi_i^*$ and \cref{eq:hamiltonian_quantised_single_reactive_jump,eq:hamiltonian_quantized_crn}, $\mathcal{S}_{DP} \left[ \left\{ \phi_{i}^*, \phi_{i} \right\} \right]$ derived for the transition dynamics reads \cite{Doi_1976,Doi_1976_2,Peliti,Peliti_1986,ATM_2024_nr_cg,Weber_2017}:
\begin{equation}\label{eq:doi_peliti_action_defination_coherant_state_picture}
\begin{split}
    \mathcal{S}_{DP} \left[ \{ \phi^*_{i}, \phi_{i} \} \right] = \int_{t_i}^{t_f} dt \mathcal{L} \left[ \{ \phi^*_{i}, \phi_i \} \right],
\end{split}
\end{equation}
with the mesoscopic Lagrangian $\mathcal{L} \left[ \{ \phi^*_{i}, \phi_i \} \right]$ and the mesoscopic Hamiltonian $\mathcal{H}[\{\phi_i^*, \phi_i\}]$ are,
\begin{equation}\label{eq:Lagrangian_coherant_mesoscopic}
\begin{split}
    \mathcal{L} \left[ \{ \phi^*_{i}, \phi_i \} \right] & = - \sum_{ \{i\} } \phi_i \partial_t \phi_i^* - \mathcal{H}[\{\phi_i^*, \phi_i\}],
    \\
    \mathcal{H}[\{\phi_i^*, \phi_i\}] & = - \sum_{\{\gamma^\rightleftharpoons \}} \left( \phi_{r_\gamma}^* - \phi_{p_\gamma}^* \right) \left( \phi_{r_\gamma} k_\gamma - \phi_{p_\gamma} k_{-\gamma} \right)
\end{split}
\end{equation}
In the definition of $\mathcal{L} \left[ \{ \phi_{i}, \phi_i^* \} \right]$, the first and second terms lie in the state-space and transition-space, respectively. Their physical origin is attributed to the left and right sides of the Master \cref{eq:master_equation}. Excluding the boundary terms, the first term of $\mathcal{L} \left[ \{ \phi^*_{i}, \phi_i \} \right]$ is reorganized and equal to $ \sum_{\{i\}} \phi_i^* \partial_t \phi_i$. 
\vspace{3pt}

\textit{Cole-Hopf transform and mesoscopic Doi-Peliti Lagrangian in the density-affinity picture}. \textemdash \:
\Cref{eq:doi_peliti_path_integral_coherant_state_picture,eq:doi_peliti_action_defination_coherant_state_picture,eq:Lagrangian_coherant_mesoscopic} formulate the `stochastic' path integral representation of the discrete-state process in the second-quantized coherent-state picture. However, the coherant state eigenvalues $\phi_i$ and $\phi_i^*$ do not have intuitive physical meaning for classical stochastic systems \cite{itakura_2010,ATM_2024_nr_cg}. The Cole-Hopf transform transforms the complex eigenvalue fields $\phi_i$ and $\phi_i^*$ to classical fields: the state probability/density ($\rho_i$) and the corresponding conjugate field ($\chi_i$) \cite{itakura_2010,ATM_2024_nr_cg}. The conjugate field $\chi_i$ has been referred to by multiple names: response field, noise field, or bias field. However, we will stick to its mathematical nomenclature convention, namely, the conjugate field that generates state-cumulants. 

We define the Cole-Hopf transform as
$\phi_i = \rho_i e^{\chi_i}$ and $ \phi_i^* = e^{-\chi_i}$. Compared to the usual definition of Cole-Hopf transform \cite{itakura_2010}, we have defined the Cole-Hopf transform with a negative sign in $\chi_i$. This convention allows $\chi_i$ to have the same sign as the transition affinity $A_\gamma$ or the chemical potential $\mu_i$ of the state $\rho_i$. Hence, it gives a physical and thermodynamic interpretation to conjugate fields $\{\chi_i\}$, elaborated below. The Cole-Hopf transform changes the Lagrangian $\mathcal{L} \left[ \{ \phi^*_{i}, \phi_i \} \right]$ to $\mathcal{L} \left[ \{ \rho_{i}, \chi_i \} \right] = \mathcal{L} \left[ \{ \phi^*_{i}, \phi_i \} \right]|_{CF}$, in a physically intuitive and legible form, 
\begin{equation}\label{eq:Lagrangian_desnsity_affinity_mesoscopic}
\begin{split}
    \mathcal{L} \left[ \{ \rho_{i}; \chi_i \} \right] 
    & = \vec{\chi} \cdot \partial_t \vec{\rho} - \mathcal{H}[\{\rho_i, \chi_i\}], 
    \\
    \mathcal{H}[\{\rho_i; \chi_i\}] 
    & = \sum_{\{\gamma^\rightleftharpoons\}} \big[ j_\gamma  \left( e^{\chi_{r_\gamma} - \chi_{p_\gamma}} - 1 \right) 
    + j_{-\gamma}  \left( e^{\chi_{p_\gamma} - \chi_{r_\gamma}} - 1 \right) \big]
\end{split}
\end{equation}
\subsubsection{Doi-Peliti Lagrangian in the current-affinity picture}\label{sec:var_dp_transition_space}
The first and second terms of $\mathcal{L} \left[ \{ \rho_{i}; \chi_i \} \right]$ lie in the state-space and the transition-space, respectively, similar to \cref{eq:doi_peliti_action_defination_coherant_state_picture}. Hence, our objective is to represent the Lagrangian $\mathcal{L} \left[ \{ \rho_{i}; \chi_i \} \right]$ in the transition space, since the microscopic transitions are the most fundamental physical origin of the system's stochasticity. To this end, the first term of $\mathcal{L} \left[ \{ \rho_{i}; \chi_i \} \right]$ is rewritten using the continuity \cref{eq:master_equation}, which equals replacing $\partial_t \rho_i$ by $j_\gamma$ for all transitions $\gamma^{\rightleftharpoons}$ that contribute to the temporal change of state $\rho_i$ due to the continuity equation. State-space conjugate fields are also represented in the transition space by defining $\chi_\gamma = \chi_{p_\gamma} - \chi_{r_\gamma}$ and $\chi_{-\gamma} = \chi_{r_\gamma} - \chi_{p_\gamma}$, hence, by definition, $\chi_{\gamma}$ satisfies the topological constraint $\chi_{-\gamma} = - \chi_{\gamma}$. 

This reduces the state-space representation of the Lagrangian $\mathcal{L}[\{\rho_i; \chi_i\}]$ to the transition-space representation $\mathcal{L}[\{j_\gamma; \chi_\gamma\}]$ for each unidirectional transition $\gamma^\rightharpoonup$.
\begin{equation}\label{eq:doi_peliti_transition_lagrangian_unidirectional}
\begin{split}
    \mathcal{L}[\{ j_\gamma; \chi_\gamma \}] = \sum_{\{ \gamma^\rightharpoonup \}} j_\gamma \left( \chi_\gamma + 1 - e^{-\chi_\gamma} \right).
\end{split}    
\end{equation}
Furthermore, recalling definitions $\chi_{-\gamma} = - \chi_{\gamma}$ and $J_\gamma = j_\gamma - j_{-\gamma}$ and $T_\gamma = j_\gamma + j_{-\gamma}$, \cref{eq:doi_peliti_transition_lagrangian_unidirectional} is simplified to:
\begin{equation}\label{eq:doi_peliti_transition_lagrangian}
\begin{split}
    \mathcal{L}  \left[ \{ J_\gamma, T_\gamma; \chi_\gamma \} \right]
    & = \sum_{\{\gamma^{\rightleftharpoons}\}} 
    \left[ 
    J_{\gamma} \left( \chi_\gamma + \sinh{ \left( \chi_\gamma \right) } \right) + T_{\gamma} 
    \left( 1 - \cosh{ \left( \chi_\gamma \right) }\right)
    \right],
\end{split}    
\end{equation}
defined for the set of all bidirectional transitions.
\Cref{eq:doi_peliti_transition_lagrangian} incorporates time-reversal symmetry/asymmetry due to the topological constraint imposed on $\chi_\gamma$. It reveals that the symmetric and anti-symmetric parts of transition currents are coupled to a single conjugate field $\chi_\gamma$.  Moreover, the time anti-symmetric term $J_\gamma$ and the time-symmetric term $T_\gamma$ are coupled to odd and even powers of $\chi_\gamma$, respectively. This structure of \cref{eq:doi_peliti_transition_lagrangian} reveals that the statistical properties of all higher-order transition fluctuations are effectively encapsulated by the first and second cumulants ($J_\gamma$ and $T_\gamma$), due to the nonlinear coupling to the conjugate field $\chi_\gamma$.

Combining \cref{eq:doi_peliti_path_integral_coherant_state_picture,eq:doi_peliti_action_defination_coherant_state_picture} with \cref{eq:doi_peliti_transition_lagrangian}, the `stochastic' path integral representation of the discrete-state processes in the current-affinity picture is given by the transition probability measure,
\begin{equation}\label{eq:doi_peliti_path_integral_current_affinity}
\begin{split}
    \mathcal{P} \left[ \left\{ J_{\gamma}, T_\gamma ; \chi_{\gamma} \right\} \right]
    = e^{ -\mathcal{S}_{DP} \left[ \left\{ J_{\gamma}, T_\gamma; \chi_{\gamma} \right\} \right] },
\end{split}
\end{equation}
and the corresponding Doi-Peliti action and Lagrangian:
\begin{equation}\label{eq:doi_peliti_action_defination_density_affinity}
\begin{split}
    \mathcal{S}_{DP} \left[ \{ J_{\gamma}, T_\gamma; \chi_{\gamma} \} \right] = \int_{t_i}^{t_f} dt \mathcal{L} \left[ \{ J_{\gamma}, T_\gamma; \chi_{\gamma} \} \right],
\end{split}
\end{equation}
The coherent-state sums over all Poissonian realization of the state, which is encapsulated in its eigenvalues, but the stochasticity associated with the microscopic stochastic transition between states is preserved through the conjugate fields $\{ \chi_\gamma \}$, which couples non-linearly to both $J_\gamma$ and $T_\gamma$, and the transition probability measure is quantified by \cref{eq:doi_peliti_path_integral_current_affinity}. The Stratonovich-Hubbard transform is an example of such path-integral formulism, where a conjugate field $\chi_{o}$ couples linearly to an observable $\mathcal{O}$, and quantifies it's gaussian fluctuations (stochasticity of the observable $\mathcal{O}$) with a quadratic coupling to $\chi_{o}$, the validity of this approach relies on cEQ Gaussian approximation: \textit{a top-down approach} towards the fluctuation of the relevant coarse-grained observable \cite{Onsager_1953,Onsager_1953_2}, detailed discussion in \cref{sec:gaussian_rate_functional}. However, the non-linear coupling of $\chi_\gamma$ to $J_\gamma$ and $T_\gamma$ in \cref{eq:doi_peliti_transition_lagrangian_unidirectional,eq:doi_peliti_transition_lagrangian} is attributed to incorporating Poissonian transitions between states, due to the exact \textit{bottom-up approach} towards the transition fluctuations. The set of \cref{eq:doi_peliti_transition_lagrangian,eq:doi_peliti_path_integral_current_affinity,eq:doi_peliti_action_defination_density_affinity} concludes the derivation of the exact `stochastic' path integral formulism for discrete-state processes \cite{Weber_2017}, which incorporates poissonian transition fluctuations and is the central result of this work.

{
The results obtained in \cref{sec:var_dp_theory} fall within the existing DPFT framework, which is formulated in state space. In contrast, the transition-space representation developed in \cref{sec:var_dp_transition_space} constitutes the main technical advance of this work. Since $\chi_i$ is the generator associated with the state $\rho_i$, its state-space nature prevents it from tracking the specific transition channel through which a state change occurs. By comparison, the transition-space representation in \cref{eq:Lagrangian_desnsity_affinity_mesoscopic} assigns a generator $\chi_\gamma$ to each linearly independent transition. To illustrate this distinction, consider again the example of two protein conformations undergoing transitions via two channels: an equilibrium channel and an enzymatically driven one. Using $\chi_{r_{eq}} = \chi_1$, $\chi_{p_{eq}} = \chi_2$, $\chi_{r_{enz}} = \chi_1$, and $\chi_{p_{enz}} = \chi_2$, \cref{eq:Lagrangian_desnsity_affinity_mesoscopic} reduces to a function of $\chi_1 - \chi_2$, i.e., a state-space representation that cannot distinguish which transition channel is taken. In contrast, \cref{eq:doi_peliti_transition_lagrangian}, formulated in transition space, depends on $\chi_{eq}$ and $\chi_{enz}$, and therefore explicitly resolves the contribution of each transition channel. This also highlights the physical equivalence of the conjugate fields $\chi_i$ and $\chi_\gamma$, which can be interpreted, respectively, as a stochastic chemical potential associated with the state and a stochastic affinity associated with the transition, discussed below.
}
%
%
%
%
\subsubsection{Most likelihood path and the inferred EPR}\label{sec:most_likelihood_path}
\begin{figure}[t!]
\centering
\includegraphics[width=\linewidth]{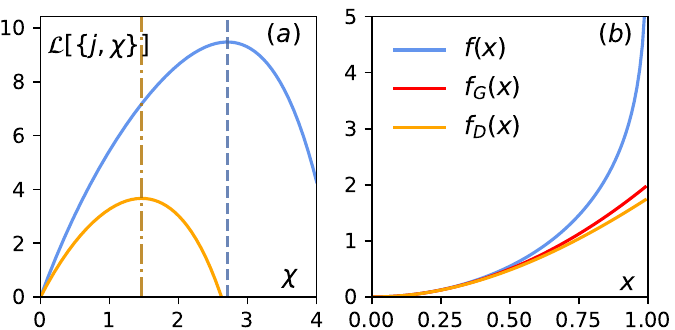}
\caption{ (a) Lagrangian $\mathcal{L}[j, \chi]$ for fixed $J_\gamma=3.5, T_\gamma=4$ (cyan) and $J_\gamma=2.5, T_\gamma=4$ (orange). The corresponding most-likely transition affinity $\chi^* = 2\tanh^{-1}(J_\gamma/T_\gamma)$ is shown as a vertical dotted line. (b) Comparison between the exact large deviation rate functional $I = 2x\tanh^{-1}(x)$, the dynamical rate functional $I_{D}= 2x\sinh^{-1}(x)$, and the close-to-equilibrium quadratic (Gaussian) approximated rate functional $I_{G} = 2x^2$, where $x=J_\gamma/T_\gamma$ is the current precision. }
\label{fig:1}
\end{figure}

The transition probability measure \cref{eq:doi_peliti_path_integral_current_affinity} is dominated by the saddle-point of \cref{eq:doi_peliti_transition_lagrangian}. Therefore, the most-likelihood path gives the optimization problem for the Lagrangian with respect to the conjugate affinity $\chi_\gamma$:
\begin{equation}\label{eq:variational_principle_EPR}
\begin{split}
    \mathcal{L}^*  \left[ \{ J_\gamma, T_\gamma \} \right] 
    = \sup_{ \{\chi_\gamma \} } \mathcal{L}  \left[ \{ J_\gamma, T_\gamma; \chi_\gamma \} \right].
\end{split}
\end{equation}
The extremization $\delta \mathcal{L}  \left[ \{ J_\gamma, T_\gamma; \chi_\gamma \} \right] / \delta \chi_{\gamma} = 0$ gives the optimal `effective' affinity for the stochastic current $\chi_{\gamma}^* = 2 \tanh^{-1}{\left(J_{\gamma} / T_{\gamma} \right)}$ and the corresponding `effective' Lagrangian $\mathcal{L}^*  \left[ \{ J_\gamma, T_\gamma \} \right]$,
{
where, for any given stochastic realization of $J_\gamma$ and $T_\gamma$, $\chi_\gamma^*$ and $\mathcal{L}^*$ are unique.
}
Physically, it corresponds to the most likelihood transition affinity that generates the given instantaneous stochastic current and traffic;
see \cref{fig:1}\textcolor{red}{(a)}. It will play a key role throughout this work. 

The saddle-point approximation parameter is $1$ for the extremization with respect to $\chi_\gamma$. This physically corresponds to incorporating Poisssonian transition fluctuations for mesoscopic systems, with a quantitatively equal weight to all higher-order current cumulants. The extremization procedure in \cref{eq:variational_principle_EPR} is mathematically equivalent to the inverse of the Stratonovich-Hubbard transform \cite{Stratonovich_1957,Hubbard_1959}, with a non-linear coupling to the conjugate field. Due to which, it leads to a non-quadratic dependence of $\mathcal{L}^*$ on $J_\gamma$ and $T_\gamma$, because of the quantification of non-Gaussian fluctuations, due to the `\textit{bottom-up construction}' towards the transition fluctuations developed here. In contrast, the usual quadratic formulation corresponds to Gaussian fluctuations and yields a quadratic Onsager-Machlup functional \cite{Onsager_1953,Onsager_1953_2}. This assumes the validity of the mean-field approximation of the transition currents and incorporates the gaussian fluctuations around it by implementing the Stratonovich-Hubbard transform that couples the conjugate field to the current and traffic linearly and quadratically, respectively; see \cref{eq:lagrangian_gaussian}: a \textit{top-down approach} towards incorporating the transition fluctuations \cite{Martin_1973,Bausch_1976,Janssen_1976,Dedominicis_1976}.

$\chi_{\gamma}^*$ depends non-linearly on the precision of the current defined as the ratio $J_\gamma / T_\gamma$. Physically, this is a consequence of exactly incorporating the non-Gaussian transition fluctuations. Therefore, taking into account all higher-order cumulants renormalizes the effective affinity to give a non-linear relation between the current precision and the effective affinity. Remarkably, the first two cumulants completely determine the exact transition statistics, a physical manifestation of the $\mathbb{Z}_2$ symmetry of the forward and backward transitions, a consequence of time-reversal symmetry imposed by the transition topology. Then, the `effective' Lagrangian $ \mathcal{L}^*[\{ J_{\gamma}, T_{\gamma} \}]$ reads: 
\begin{equation}\label{eq:onsager_Machlup_functional}
\begin{split}
    \dot{\Sigma} = \mathcal{L}^*[\{ J_{\gamma}, T_{\gamma} \}]   
    & = \sum_{ \{ \gamma^{\rightleftharpoons} \} } 2 J_\gamma \tanh^{-1}{ \left( \frac{J_\gamma}{T_\gamma} \right) },
\end{split}
\end{equation}
If the transition affinities have been known, \cref{eq:onsager_Machlup_functional} is equal to the mean transition EPR. This follows trivially using analytical expressions $ J_\gamma = 2D_\gamma \sinh{(A_\gamma/2)}$ and $ T_\gamma = 2D_\gamma \cosh{(A_\gamma/2)}$ which implies $\chi_\gamma^* = A_\gamma$. However, \cref{eq:onsager_Machlup_functional} defines EPR for a given stochastic realization of $J_\gamma$ and $T_\gamma$, without knowing the transition affinities, which is equivalently seen by the bilinear form of \cref{eq:onsager_Machlup_functional}, $\mathcal{L}^*[\{ J_{\gamma}, T_{\gamma} \}] =  \sum_{ \{ \gamma^{\rightleftharpoons} \} } J_\gamma \chi_\gamma^*$ and gives a thermodynamic meaning to $ \mathcal{L}^*[\{ J_{\gamma}, T_{\gamma} \}]$: the `inferred' mean EPR in the absence of knowledge about the transition affinities.  
 
$\chi_\gamma^* = A_\gamma$ does not necessarily hold. To highlight this point, consider that an observed value of current and traffic are $J_\gamma = 2.5$ and $T_\gamma = 4$, which results in an effective inferred affinity $\chi_\gamma^*$, vertical dark orange dashed line in \cref{fig:1}\textcolor{red}{(a)}, the corresponding underlying effective Lagrangian that generates this most likelihood path is given by the solid orange curvy line in \cref{fig:1}\textcolor{red}{(a)}, this case corresponds to $\chi_\gamma^* = A_\gamma$. However, this realization of observed stochastic current and traffic could have been a lesser likelihood path of a different underlying model, for example, the intersection point of the cyan solid line (other model) and the vertical dark orange dashed line, this case corresponds to $\chi_\gamma^* \neq A_\gamma$, since $A_\gamma$ is given by the dotted vertical dark cyan line in \cref{fig:1}\textcolor{red}{(a)}. This allows defining the inferred mean EPR for any given lesser likelihood stochastic realization of the current and traffic, without knowing the underlying affinities. The inferred mean EPR is equal to the deterministic mean EPR of the underlying true model, only if the sampled realization of the stochastic current and traffic is the most likelihood path, otherwise it is a lesser likelihood fluctuation of EPR.

\subsubsection{ Thermodynamic length and entropy production }\label{sec:tl_ep}
Defining a time-integrated stochastic microscopic current $ \tau \tilde{J}_{\gamma} = \int_{0}^{\tau} \hat{J}_{\gamma} $ and traffic $ \tau \tilde{T}_{\gamma} = \int_{0}^{\tau} \hat{T}_{\gamma} $. Here, $\int_0^\tau$ is defined using the counting observable such that $\tau \tilde{J}_{\gamma}$ and $\tau \tilde{T}_{\gamma}$ count the total directional current and bidirectional transitions for $\gamma^\rightleftharpoons$, respectively. Physically, $\tau \tilde{J}_{\gamma}$ and $\tau \tilde{T}_{\gamma}$ quantify the thermodynamic length and dynamical activity of the transition $\gamma^\rightleftharpoons$ over the given observation time $\tau$. Here, $\tilde{J}_\gamma$ and $\tilde{T}_\gamma$ are scaled time-integrated stochastic current and traffic, with the observation time $\tau$ being the scaling parameter that assumes the dissipative scaling of the transition currents. We integrate \cref{eq:onsager_Machlup_functional} from the initial time $t=0$ to the final time $\tau$ and obtain the following non-quadratic relation between the thermodynamic length and the EP,
\begin{equation}\label{eq:EP_and_thermodynamic_length}
\begin{split}   
    \tau \tilde{\Sigma} = \Sigma = \mathcal{S}_{DP}^* = \int_0^{\tau} \mathcal{L}^* d t & \geq \sum_{ \{\gamma^{\rightleftharpoons}\} } 2 \tau \tilde{J}_{\gamma} \tanh^{-1}{ \left( \frac{ \tilde{J}_{\gamma} }{ \tilde{T}_{\gamma} } \right) },
\end{split}
\end{equation}
where, we have used Jensen's inequality to integrate \cref{eq:onsager_Machlup_functional}, which turns the equality between $\dot{\Sigma}$ and $J_\gamma, T_{\gamma}$ in \cref{eq:onsager_Machlup_functional} to the inequality (lower bound) between ${\Sigma}$ and $\tilde{J}_\gamma, \tilde{T}_{\gamma}$ in \cref{eq:EP_and_thermodynamic_length}, such that the equality is recovered in $\tau \to 0$. 

\Cref{eq:onsager_Machlup_functional,eq:EP_and_thermodynamic_length} relate the short-time and finite-time non-quadratic thermodynamic lengths to the thermodynamic quantities $\dot{\Sigma}$ and $\Sigma$. It reveals that the dynamic quantities ( $\tilde{J}_\gamma/J_\gamma$ and $\tilde{T}_\gamma/T_\gamma$ here) of the fEQ systems are fundamentally constrained by the thermodynamic EP/EPR, and delineates a trade-off between current mean, current fluctuations, and dissipation. The exact formulation of the non-quadratic thermodynamic length gives the tightest bound compared to the known phenomenological quadratic and non-quadratic counterparts. In contrast, equilibrium or cEQ systems exhibit a quadratic relation of thermodynamic length and entropy production and are valid for the free energy or excess EP, respectively \cite{Salamon_1983,Salamon_1985,Schlogl_1985,Crooks_2007,Ito_2018}, which represents only a part of the total EP, reducing its applicability to fEQ systems due to massive underestimation of EPR. 
\subsubsection{Min-Max principle for the EPR }\label{sec:min_max_formulation}
The set of \cref{eq:doi_peliti_transition_lagrangian_unidirectional,eq:doi_peliti_path_integral_current_affinity,eq:doi_peliti_transition_lagrangian,eq:variational_principle_EPR,eq:onsager_Machlup_functional,eq:doi_peliti_action_defination_density_affinity}, formulates a min-max variational problem for the action, in particular, $ \Sigma = \inf_{\{J_{\gamma}, T_{\gamma} \}} \sup_{ \{\chi_{\gamma} \} } \mathcal{S}_{DP} \left[ \left\{ J_{\gamma}, T_{\gamma}; \chi_{\gamma} \right\} \right] $. It is a minimum action principle valid for fEQ systems, which allows a unified description of stochastic discrete state systems \cite{atm_2024_var_epr}. The min-max principle is a variational consequence of \cref{eq:doi_peliti_transition_lagrangian} being concave in $\chi_\gamma$ (\cref{fig:1}\textcolor{red}{(a)}), and \cref{eq:onsager_Machlup_functional} being convex in $J_\gamma/T_\gamma$ (blue line in \cref{fig:1}\textcolor{red}{(b)}) combined with the saddle-point approximation that aims to minimize the action. The min-max formulation here is more subtle than the existing cEQ formulations. It requires a first maximization over the conjugate affinity and then a subsequent minimization over the transition currents. The maximization over the conjugate affinity is more important for systems exhibiting prominent stochastic effects. To elaborate on this point, we highlight two major fundamental regimes for the applicability of min-max principle that lead to different physical principles. 

The first regime corresponds to the constraining of the effective transition affinity $\chi_\gamma$, defined as, `force constraint' systems. In this regime, min-max formulation is effectively realized as the `minimization of inferred EPR', since $\inf_{\{J_{\gamma}, T_{\gamma} \}} \sup_{ \{\chi_{\gamma} \} } \mathcal{S}_{DP} \left[ \left\{ J_{\gamma}, T_{\gamma}; \chi_{\gamma} \right\} \right] = \inf_{\{J_{\gamma}, T_{\gamma} \}} \mathcal{S}_{DP} \left[ \left\{ J_{\gamma}, T_\gamma \right\} \right]$. The `force constraint' system is physically realized in two important scenarios: when $\chi_\gamma$ is constant and/or small. First, when the noise effects (stochasticity) are not prominent, the effective transition affinity is a constant, for example, a deterministic limit of chemical reaction networks, and the study of non-equilibrium steady-state analysis using irreversible thermodynamics \cite{Glansdorff_1971}. These systems have been paradigms for justifying the `minimum entropy production principle' \cite{Glansdorff_1971}. Second, when $\chi_\gamma$ is small, it quantifies cEQ gaussian fluctuations using the quadratic Onsager-Machlup functional \cite{Onsager_1953,Onsager_1953_2}. The quadratic Onsager-Machlup functional has been used to study fluctuations of fEQ systems, despite it being obtained relying on the cEQ Gaussian approximation.

The second regime corresponds to constraining the transition current $J_\gamma$, namely `current constraint' systems. In this case, min-max formulation is effectively realized as the maximization of inferred EPR namely MaxEP, because $\inf_{\{J_{\gamma}, T_{\gamma} \}} \sup_{ \{\chi_{\gamma} \} } \mathcal{S}_{DP} \left[ \left\{ J_{\gamma}, T_{\gamma}; \chi_{\gamma} \right\} \right] = \sup_{ \{\chi_{\gamma} \} } \mathcal{S}_{DP} \left[ \left\{ J_{\gamma}, T_{\gamma}; \chi_{\gamma} \right\} \right]$. Hence, physically, this implies that, if the transition currents are fixed using an external thermodynamic reservoir, the system maximizes the corresponding effective affinity, which effectively minimize the corresponding current fluctuations. Equivalently, a similar mechanism applies to systems prone to non-Gaussian stochastic effects. Here, maximization over the conjugate noise field becomes important, and hence the system effectively realizes 'MaxEP' \cite{Endres_2017}. Physically, `MinEP' and `MaxEP' are valid approximations of the Min-Max principle. 

{
However, more practical systems lie in an intermediate regime, where partially `current constraint' and `force constraint' on the system can coexist. Such sophisticated setups are expected to exhibit a more complex interplay between the `minimization' and `maximization' of entropy production, with the transition between them dictated by the control parameters of the specific model studied \cite{Endres_2017}. Nevertheless, the MinAP is expected to hold for any control parameter regime. In this respect, similar to equilibrium systems, MinAP can incorporate experimental observable constraints via a Lagrange multiplier.
}

Defining conjugate affinities $\chi_{J_\gamma} = \partial_{J_\gamma} \mathcal{L}^*$ and $\chi_{T_\gamma} = \partial_{T_\gamma} \mathcal{L}^*$ for currents and traffics, $\mathcal{L}^*[\{J_\gamma,T_\gamma\}]$ satisfies the zero-cost flow constraint $\sup_{\{J_{\gamma},T_{\gamma}\}} \left( J_{\gamma} \chi_{J_\gamma} + T_{\gamma} \chi_{T_\gamma} - \mathcal{L}^* \right) = 0 $, establishing the connection between the mean inferred EPR and frenetic activity \cite{Maes_2017}. Mathematically, this dependence was rather clear from \cref{eq:doi_peliti_transition_lagrangian}, where the conjugate field $\chi_\gamma$ couples to both $J_\gamma$ and $T_\gamma$, but we still explicitly highlight it.
\subsubsection{Mapping to information geometry}\label{sec:mapping_ig}
From an information geometric viewpoint, the right-side of \cref{eq:variational_principle_EPR} is the variational representation of the KL divergence defined between the forward and backwards path probability measures, which is defined as the EPR in \cite{kolchinsky_2022_information_geometry_epr,Kolchinsky_2024,Amari_2016_book} and other relevant results on variational formulation \cite{Otsubo_2020,kim_2020,Otsubo_2022}. This is prominently visible using the representation of Lagrangian in unidirectional transition currents \cref{eq:doi_peliti_transition_lagrangian_unidirectional}, which is the Donsker-Varadhan representation of KL divergence used in Information geometry, and is equal to $\sup_{\{\chi_\gamma\}} \mathcal{L}[\{j_\gamma; \chi_\gamma\}]$ \cite{kolchinsky_2022_information_geometry_epr,Kolchinsky_2024,Amari_2016_book}. Hence, our formulation proves the equivalence of the EPR obtained using the information geometric and statistical mechanical formulation. Importantly, in contrast to Ref.\cite{kolchinsky_2022_information_geometry_epr,Kolchinsky_2024,Amari_2016_book}, we have defined EPR using the microscopic observables of the model itself and does not require identification of the backward process \footnote{Due to our \textit{bottom-up approach}, we had identified the backward process on a microscopic level by decomposing the unidirectional transitional currents into its time-antisymmetric ($J_\gamma$) and time-symmetric ($T_\gamma$) components.}, which is sensitive to correct/incorrect identification of the backward process, due to the resolution of the given trajectory/path. This exact mapping allows us to study ST using information geometry as a mathematical tool with a physical interpretation given by the `Minimum action principle'.

\subsubsection{Beyond the local detailed balance condition}
\label{sec:beyond_LDB}

{
$\chi_\gamma$ is a parameter that is associated with the statistical properties of the system. In comparison, $A_\gamma$ is associated with the thermodynamically-consistent model definition imposed through the LDB condition required in stochastic thermodynamics. However, the derivation of MinAP does not utilize LDB condition. Therefore,
the `stochastic' path integral formulation and `MinAP' derived in \cref{sec:var_formulation} and its physical formulation as a fEQ analogue of the Boltzmann distribution holds irrespective of the LDB condition for a broad class of fEQ discrete-state systems beyond validity regimes of Stochastic Thermodynamics. When the LDB condition is satisfied, the thermodynamic consistency of transition currents endows the effective Lagrangian with a clear physical interpretation as the mean EPR, as discussed in \cref{sec:most_likelihood_path}. Even in the absence of LDB, the information-geometric mathematical interpretation of the Lagrangian remains valid. Importantly, even if the LDB condition is violated, or if the `observer' is practically agnostic about its validity (even when it holds microscopically), the `physical' inferred mean EPR, which is constructed from the time-symmetric and time-antisymmetric components of physical observables (such as currents and traffics), remains well-defined. Owing to this distinction, the information-geometric interpretation presented here is directly grounded in physically observable quantities.
}

{
$\chi_\gamma^*$ being time-homogeneous (independent of observation time) is equivalent to the system being memory-less (if valid $\forall \: \gamma^{\rightleftharpoons} \in \{ \gamma^\rightleftharpoons\}$); therefore, in this regime, due to the time-scale separation, the initial-value condition for the states and the correlations between the system and the thermodynamic reservoir become irrelevant, and the `\textit{bottom-up}' constructed validity of the LDB condition emerges naturally \cite{Evans_2004,Evans_2005}.
Since the derivation of MinAP does not rely on the LDB condition, it allows the study of transient effects that violate LDB and goes beyond the assumption of timescale separation in ST. In ST, the weak coupling between the system and the reservoir allows for the formulation of the LDB condition, which in itself is interpreted as a form of cEQ assumption (in the sense of weak coupling between the system and environment). This highlights that the MinAP extends the dual thermodynamic structure between affinities and the currents and traffics to fEQ systems, beyond the weak coupling regime in ST.
}
\subsection{Large deviation principle}\label{sec:ldp}
The Boltzmann distribution establishes an equivalence between the statistical properties of the physical quantities and the energetics of the systems \cite{boltzmann_1964}. Using the variational formulation and the large deviation theory, our aim is to investigate a similar principle for fEQ systems \cite{Touchette_2009}. The large deviation theory studies fluctuations of dynamical observables in non-equilibrium systems \cite{Touchette_2009}. The probability distribution $\mathcal{P}(\mathcal{O})$ for non-equilibrium physical observables $(\mathcal{O})$ is said to satisfy the large deviation principle with a large deviation parameter $\Omega$ and a rate functional $I(\mathcal{O})$, if it satisfies $\mathcal{P}(\mathcal{O}) \asymp e^{-\Omega I(\mathcal{O})}$. The scaling parameter $\Omega$ dictates the convergence of the probability distribution to the minimum of the rate functional (the most likelihood value of the observable) and also quantifies the fluctuations around the most likelihood value. Here, the observable is an intensive variable and satisfies the scaling for the mean $\langle \mathcal{O} \rangle \propto O(1)$ and variance $Var(\mathcal{O}) \propto 1/ \Omega$. The Boltzmann distribution for equilibrium systems $\left( \mathcal{P}^{eq} \asymp e^{ - \beta E} \right)$ is an example of LDP. Where, the inverse temperature $\beta$ is the LDP scaling parameter, and the equilibrium energy functional $E$ is the corresponding rate functional that quantifies the energy of the system. Fluctuations vanish in the zero-temperature limit, and the system's energy converges to the minimum of $E$.

However, the LDP formulation is prone to two assumptions/approximations. (1): the existence and particular analytical form of such a rate functional. For instance, a Gaussian approximation for the observable statistics. (2): Choosing a coarse-grained macroscopic observable that discards information about other relevant microscopic observables. Here, we aim to address both issues and derive an exact rate functional using a systematic \textit{`bottom-up approach'} for all microscopically relevant physical quantities. To highlight the importance of the exact LDP, we will compare it to two phenomenological cases that correspond to the Gaussian approximation of noise and dynamical rate functionals, which correspond to the thermodynamic uncertainty relation and the non-equilibrium fluctuation-response relation, respectively.
\subsubsection{The exact result}\label{sec:exact_rate_functional}
We combine \cref{eq:EP_and_thermodynamic_length,eq:doi_peliti_path_integral_current_affinity} and obtain the exact LDP for the scaled-time-integrated current and traffic, $\tilde{J}_\gamma$ and $\tilde{T}_\gamma$, it reads,
\begin{equation}\label{eq:probability_exact_rate_functional}
\begin{split}
    \mathcal{P} \left[ \left\{ \tilde{J}_{\gamma}, \tilde{T}_{\gamma} \right\} \right] \asymp
    e^{-\tau I \left[ \left\{ { \tilde{J}_{\gamma} }, { \tilde{T}_{\gamma} } \right\} \right] },
\end{split}    
\end{equation}
where, $I\left[ \tilde{J}_\gamma, \tilde{T}_\gamma \right] = 2 \tilde{J}_\gamma \tanh^{-1}{\left( {\tilde{J}_\gamma}/{\tilde{T}_\gamma}\right)}$ is the exact dynamical rate functional for $\tilde{J}_\gamma$ and $\tilde{T}_\gamma$, with the observation time $\tau$ is the corresponding LDP scaling parameter for the dynamical canonical ensemble \cite{Touchette_2009,Maes_2017}. Due to the second saddle-point approximation (with respect to $\tau$), the transition probability measure for $\tilde{J}_\gamma$ and $\tilde{T}_\gamma$ peaks at the minimum of the rate functional \cite{Touchette_2009,Maes_2017}. 

We define the current precision for the finite- and short-time processes $\tilde{x}_\gamma = \tilde{J_\gamma}/\tilde{T_\gamma}$ and ${x}_\gamma = {J_\gamma}/{T_\gamma}$, respectively. $I[\tilde{J}_\gamma, \tilde{T}_\gamma]$ is rewritten using the current precision and $f(x) = 2x \tanh^{-1}{(x)}$ such that $ I \left[  { \tilde{J}_{\gamma} }, { \tilde{T}_{\gamma} } \right] = \tilde{T}_{\gamma} f \left( \tilde{x}_{\gamma}  \right) $. This scaling implies that $\tilde{T}_\gamma$ defines the timescale corresponding to $\gamma^\rightleftharpoons$, and $\tilde{x}_\gamma$ is the relevant observable with rate funtional $f(\tilde{x})$. We can absorb $\tilde{T}_\gamma$ by redefining time $t' = t \tilde{T}_\gamma$. However, different transitions have different timescales, so $\tilde{T}_\gamma$ is also a relevant parameter. Importantly, because the scaling parameter $\tau$ characterizes the convergence of the transition probability measure to the minimum of the rate functional (most likelihood value), with a relaxation timescale $\tau_{\gamma}^{rel} = 1/\tilde{T}_\gamma$. Hence, for a given fixed observation time $\tau$, higher values of $\tilde{T}_\gamma$ accelerate the convergence of $\tilde{x}_\gamma$ to it's most likelihood value. Physically, a faster current dynamics (degree of freedom) is approximated with a constant value, and the stochasticity of slower dynamics could be studied. This physical mechanism usually comes under different names, for example, time-scale separation, adiabatic approximation of fast dynamics. 

$f(x)$ also relates the precision of time-integrated currents to the EP $\Sigma_\gamma$ associated with transition $\gamma^{\rightleftharpoons}$. Due to \cref{eq:EP_and_thermodynamic_length}, we obtain $\Sigma_\gamma = \tau \tilde{T}_\gamma f({\tilde{x}_\gamma})$. Using the inverse function $f^{-1}(x)$, we obtain the nonquadratic upper bound on the current precision for the given EP and traffic, which reads, 
\begin{equation}\label{eq:precision_bound}
\begin{split}
    \tilde{J}_{\gamma} \leq \tilde{T}_{\gamma} f^{-1}\left( \frac{\Sigma_{\gamma}}{\tau \tilde{T}_{\gamma}} \right).
\end{split}    
\end{equation}
Importantly, the scaling form of ${\Sigma_{\gamma}}/{\tau \tilde{T}_{\gamma}}$ in \cref{eq:precision_bound} signifies that the EP for a transition ${\Sigma_{\gamma}}$ should be measured in its natural timescale $\tau \tilde{T}_{\gamma}$. \Cref{eq:precision_bound} is the fundamental universal scaling relation between the precision of the time-integrated current and EP, that bounds the precision of the transition current with the scaled EP (in the natural timescale of transition). 
\Cref{eq:onsager_Machlup_functional,eq:EP_and_thermodynamic_length,eq:probability_exact_rate_functional} formulate the fundamental foundation of this work, namely an exact canonical ensemble analog for the statistical properties of dynamical physical quantities (current and traffic) and their connection to thermodynamic dissipation. 
\subsubsection{The Gaussian approximation and the Thermodynamic Uncertainty Relation}\label{sec:gaussian_rate_functional}
The Gaussian approximation of the transition fluctuations is equivalent to the second-order Taylor-series expansion of \cref{eq:doi_peliti_transition_lagrangian} in $\chi_\gamma$, 
\begin{equation}\label{eq:lagrangian_gaussian}
     \mathcal{L}_{G}[\{ J_{\gamma}, T_\gamma; \chi_{\gamma} \} ] = \sum_{\{\gamma^\rightleftharpoons \}} 2 J_{\gamma} \chi_\gamma - \frac 1 2 T_\gamma \chi_\gamma^2,
\end{equation}
which ignores all higher order current cumulants. Having solved the variational problem for \cref{eq:lagrangian_gaussian}, $\delta \mathcal{L}_G /\delta \chi_\gamma = 0$, the effective affinity for the transition $\gamma^\rightleftharpoons$ is $\chi_\gamma^* = 2J_\gamma / T_\gamma$ corresponds to the most-likelihood path. The effective Gaussian Lagrangian is,
\begin{equation}\label{eq:lagrangian_gaussian_most_likelihood}
     \dot{\Sigma}_G = \mathcal{L}_{G}^*[\{J_\gamma, T_\gamma\}] = \sum_{\{\gamma^\rightleftharpoons\}} \frac{2 J_\gamma^2}{T_\gamma}.
\end{equation}
\Cref{eq:lagrangian_gaussian_most_likelihood} is the quadratic dissipation function originally defined for Gaussian fluctuations around the equilibrium state \cite{Onsager_1953,Onsager_1953_2}, but is generalized here for Gaussian fluctuations around any NESS . This also clarifies the nomenclature convention of the inverse Stratonovich-Hubbard transform as the transform from $\mathcal{L}$ to $\mathcal{L}^*$ \cite{Stratonovich_1957,Hubbard_1959}. The inferred EPR $\dot{\Sigma}_{G}$ using the Gaussian approximation is also known as pseudo-EPR and obtains a lower bound on $\dot{\Sigma}$ \cite{Vo_2022}. Using \cref{eq:lagrangian_gaussian_most_likelihood}, the transition probability measure satisfies LDP:
\begin{equation}\label{eq:probability_measure_gaussian}
    \mathcal{P} \left[ \left\{ \tilde{J}_{\gamma}, \tilde{T}_{\gamma} \right\} \right] \asymp e^{-\tau I_G \left[ \left\{ { \tilde{J}_{\gamma} }, { \tilde{T}_{\gamma} } \right\} \right] },
\end{equation}
with scaling relation, $I_{G} [\tilde{J}_\gamma,\tilde{T}_\gamma] = \tilde{T}_\gamma f_G(\tilde{x}_{\gamma})$, where $ f_G(\tilde{x}) = 2 \tilde{x}^2$. Compared to the exact results, $\chi_\gamma^*$ is a linear function of current precision $x_\gamma$. This led to a quadratic dissipation function in \cref{eq:lagrangian_gaussian_most_likelihood,eq:probability_measure_gaussian} due to the Gaussian approximation of the transition noise.
\subsubsection{The dynamical rate functional and the non-equilibrium fluctuation-response relation}\label{sec:frr_rate_functional}
The non-equilibrium fluctuation response relation (NFRR) between the instantaneous current and the instantaneous traffic is \cite{Vlad_1994_1,Vlad_1994_2,Vlad_1994_3,Ross_1999,ATM_2024_nr_st},
\begin{equation}
    \frac{ \partial \: J_{\gamma} }{ \partial \chi_{\gamma} } = \frac{ T_{ \gamma } }{ 2 }.
\end{equation}
We have used the conjugate field $\chi_\gamma$ instead of $A_\gamma$, allowing NFRR to be parametrically evaluated around any effective affinity $\chi_\gamma^*$. Choosing $\chi_\gamma = A_\gamma$ recovers NFRR for the given physical model. Using the non-linear dependence of $T_\gamma = \sqrt{ J_\gamma^2 + 4 D_\gamma^2 }$, the non-linear force-current relation is,
\begin{equation}\label{eq:nfrr_force_current_relation}
   J_\gamma = 2 D_\gamma \sinh{\left( \frac{\chi_\gamma}{2} \right)}.
\end{equation}
By plugging $\chi_\gamma = A_\gamma$ into \cref{eq:nfrr_force_current_relation}, the non-linear relation between the mean transition current and the affinity is recovered.

We define the dual dissipation functions $\psi_\chi$ and $\psi_J$, to further unveil the Legendre dual structure between the current and the force. To this end, we use the definition of the Legendre transform, $J_{\gamma} = \partial_{\chi_\gamma} \psi_{\chi} \left( \{ \chi_\gamma \} \right)$ and $\chi_{\gamma} = \partial_{J_\gamma} \psi_{J} \left( \{ J_\gamma \} \right)$, combined with \cref{eq:nfrr_force_current_relation}, we obtain,
\begin{equation}\label{eq:large_deviation_functional}
\begin{split}
    \psi_{\chi} \left( \{ \chi_\gamma \} \right) 
    & = \sum_{\{ \gamma^{\rightleftharpoons} \}} 4 D_\gamma \left( \cosh{\left( \frac{\chi_\gamma}{2} \right)} - 1 \right),
    \\
    \psi_{J} \left( \{J_\gamma\} \right) 
    & = \sum_{\{ \gamma^{\rightleftharpoons} \}} 2 J_\gamma \sinh^{-1}{\left( \frac{J_\gamma}{2 D_\gamma}\right)} - 2\left( \sqrt{J_\gamma^2 + 4 D_\gamma^2 } - 2D_\gamma \right),
\end{split}    
\end{equation}
where, we have imposed the constraint $\psi_F \left( 0 \right) = 0$ and $\psi_J \left ( 0 \right) = 0$, physically corresponding to the vanishing EPR for vanishing forces and currents. The dual dissipation functions characterize the EPR, the variational Lagrangian corresponding to them reads \cite{Maes_2017}:
\begin{equation}\label{eq:ldp_lagrangian}
\begin{split}
    \mathcal{L}_{D}  \left[ \{ J_\gamma; \chi_\gamma \} \right]
    & = \sum_{ \{ \gamma^{\rightleftharpoons} \} } 
    \left[ 
    \psi_{\chi} + \psi_{J}
    \right].
\end{split}    
\end{equation}
Using \cref{eq:nfrr_force_current_relation} and simplifying \cref{eq:ldp_lagrangian,eq:large_deviation_functional}, the effective Lagrangian $\mathcal{L}_{D}^* = \sup_{\{\chi_\gamma\}} \mathcal{L}_{D}  \left[ \{ J_\gamma; \chi_\gamma \} \right]$ is obtained. Where, the second term of $\psi_{J} \left( \{J_\gamma^*\} \right)$ cancels with $\psi_{\chi}\left( \{ \chi_\gamma^*\} \right)$ in \cref{eq:large_deviation_functional} and leads to:
\begin{equation}\label{eq:frr_lagrangian_effective}
    \dot{\Sigma}_{D} 
    = \mathcal{L}_{D}^*  \left[ \{ J_\gamma, D_\gamma \} \right]
    = \sum_{\{ \gamma^{\rightleftharpoons} \}} 2 J_\gamma \sinh^{-1}{\left( \frac{J_\gamma}{2 D_\gamma}\right)}.
\end{equation} 
\Cref{eq:frr_lagrangian_effective} is the previously computed dynamical large deviation rate functional for the transition currents \cite{Maes_2008,Maes_2009} and recently utilized to formulate and study the `Hessian' structure for discrete state processes in Ref. \cite{Maes_2008,Maes_2009,Maes_2017,Mielke_2014_ldp,Mielke_2017,Kaiser_2018,Renger_2021,Peletier_2022,kobayashi_2022_hessian_geometry,kobayashi_2023_information_graphs_hypergraphs,Peletier_2023,Patterson_2024,Renger_2024,Kobayashi_2022,Sughiyama_2022,Renger_2023,duong_2023,Mizohata_2024}. If a LDP were to be formulated using \cref{eq:frr_lagrangian_effective}, the transition probability measure reads:
\begin{equation}\label{eq:probability_measure_frr}
\begin{split}
    \mathcal{P} \left[ \left\{ \tilde{J}_{\gamma}, \tilde{D}_{\gamma} \right\} \right] \asymp e^{-\tau I_D \left[ \left\{ { \tilde{J}_{\gamma} }, { \tilde{D}_{\gamma} } \right\} \right] },
\end{split}    
\end{equation}
with, time-integrated mobility ($\tau \tilde{D}_\gamma = \int_0^\tau \hat{D}_\gamma$) and rate functional $I_{D}\left[ \tilde{J}_\gamma, \tilde{D}_\gamma \right] = 2 \tilde{J}_\gamma \sinh^{-1}{\left( \tilde{J}_\gamma/2\tilde{D}_\gamma \right) }$ satisfying the scaling relation $I_{D} [\tilde{J}_\gamma,\tilde{D}_\gamma] = 2 \tilde{D}_\gamma f_D(\tilde{x}_{\gamma})$ with $f_D(\tilde{x}) = 2\tilde{x} \sinh^{-1}{\left(\tilde{x} \right)}$.
\subsubsection{Physical implications of the exact LDP}
The analytical form of $f(x)$ plays a key role in the relationship between the current precision and the EPR. We plot $f(x), f_G(x)$ and $f_D(x)$ in \cref{fig:1}\textcolor{red}{(b)}, which exhibit the hierarchy of inequality,
\begin{equation}\label{eq:rate_functions_heirarchy}
    f(x) \geq f_{G}(x) \geq f_{D}(x).
\end{equation}
This hierarchy of inequality, combined with \cref{eq:EP_and_thermodynamic_length} implies that the exact rate functional computes the best bound on $\Sigma_\gamma$ for a given value of current precision. Inverting \cref{eq:rate_functions_heirarchy} and using \cref{eq:precision_bound}, we obtain the hierarchy of bounds on the current precision for the given EP corresponding to the transition, 
\begin{equation}\label{eq:precision_bound_heirarchy}
\begin{split}
    \tilde{J}_{\gamma} 
    \leq \tilde{T}_{\gamma} f^{-1}\left( \frac{\Sigma_{\gamma}}{\tau \tilde{T}_{\gamma}} \right) 
    \leq \tilde{T}_{\gamma} \sqrt{ \frac{ \Sigma_{\gamma}  }{2 \tau \tilde{T}_{\gamma} } } 
    \leq 2 \tilde{D}_{\gamma} f_D^{-1}\left( \frac{\Sigma_{\gamma}}{ 2 \tau \tilde{D}_{\gamma}} \right).
\end{split}    
\end{equation}
\Cref{eq:precision_bound_heirarchy} holds for an inverse problem, when EP corresponding to a transition is known, and the objective is to obtain the tightest bound on the corresponding current precision. The inequality between the rate functional hierarchy becomes prominent for fEQ systems, that exhibit more precise currents.  

The Gaussian approximation of the rate functional \cite{Bodineau_2004,Derrida_2007,Bertini_2015,Qian_2020} has been extensively studied to obtain bounds on EP using quadratic TKUR \cite{Barato_2015,Gingrich_2016,Horowitz_2017,Horowitz_2020,Terlizzi_2019,Van_vu_2023,Kwon_2023_TUR_unified}. Here, the exact rate functional addresses the issue of massive underestimation of EPR associated with quadratic TKUR. This mismatch is particularly pronounced for fEQ systems or those exhibiting non-Gaussian fluctuations. Similarly, $I_D$ has been utilized to study the `Hessian' dual structure between force and current \cite{Maes_2008,Maes_2009,Maes_2017,Mielke_2014_ldp,Mielke_2017,Kaiser_2018,Renger_2021,Peletier_2022,kobayashi_2022_hessian_geometry,kobayashi_2023_information_graphs_hypergraphs,Peletier_2023,Patterson_2024,Renger_2024,Kobayashi_2022,Sughiyama_2022,Renger_2023,duong_2023,Mizohata_2024}. $I$ avoids an underestimation of $\dot{\Sigma}$ and obtains the tightest bound on $\dot{\Sigma}$. The proof follows using, $I[\tilde{J}_\gamma, \tilde{T}_\gamma] > I_D[\tilde{J}_\gamma, \tilde{D}_\gamma]$, since $\tilde{T}_\gamma > 2 \tilde{D}_\gamma$ combined with $f(\tilde{x}) > f_{D}(\tilde{x})$, which implies $\dot{\Sigma} \geq \dot{\Sigma}_D$. Physically, the tightness of the bounds obtained using $I(\tilde{J}_\gamma, \tilde{T}_\gamma)$ and $f(\tilde{x})$ results from the incorporation of exact non-equilibrium current fluctuations characterized by $T_\gamma$ instead of $2D_\gamma$. Where, $2 D_\gamma$ computes the lower bound on the equilibrium current fluctuations, as $T_\gamma^{eq} = j_{\gamma}^{eq} + j_{-\gamma}^{eq} =  (j_{\gamma} + j_{-\gamma})|_{F_\gamma = 0}$ due to the inequality $T_{\gamma}^{eq} \geq 2D_{\gamma} = 2 \sqrt{j_\gamma j_{-\gamma}}$. Using $T_\gamma$ instead of $2D_\gamma$, the renormalization of the variance of the non-equilibrium currents is taken into account. In contrast, \cref{eq:large_deviation_functional} assumes a constant static equilibrium diffusivity $D_\gamma$, which is the transition mobility. Using $2D_\gamma$ instead of $T_\gamma$ reveals a violation of NFRR for fEQ systems, attributed to an underestimation of variance (fluctuations) due to $2D_\gamma$. However, as formulated above, the NFRR between the transition current response and traffic holds for fEQ systems. Due to the equivalence between the exact variational formulation and the information geometric formulation, we have a mathematical proof an open problem quoted in \cite{kolchinsky_2022_information_geometry_epr,Kolchinsky_2024}, information geometric methods obtains tighter bounds on the EPR in Ref.\cite{kolchinsky_2022_information_geometry_epr,Kolchinsky_2024}, in comparison to the Hessian structure in Ref.\cite{Maes_2008,Maes_2009,Maes_2017,Mielke_2014_ldp,Mielke_2017,Kaiser_2018,Renger_2021,Peletier_2022,kobayashi_2022_hessian_geometry,kobayashi_2023_information_graphs_hypergraphs,Peletier_2023,Patterson_2024,Renger_2024,Kobayashi_2022,Sughiyama_2022,Renger_2023,duong_2023,Mizohata_2024}.  Importantly, our analysis reveals that the shortcomings of the quadratic TKUR and Hessian structure are remedied by the exact formulation.
\section{Coarse-grained Observable Currents}\label{sec:observable_current}
Microscopic transition currents and traffics assume complete information of the system. However, experimental constraints or ignorance of the microscopic transitions restrict our access to the complete information. Experimentally, coarse-grained observable (macroscopic) currents are easily accessible. This creates a necessity to examine the possibility of the variational formulation for observable currents. In this section, we extend the applicability of the variational formulation to coarse-grained observable currents. 
{
The main results are summarized in \cref{sec:observable_current_setup}. Their derivation is detailed in \cref{sec:derivation_cg_observable} for completeness. The reader is encouraged to read through the derivation \cref{sec:derivation_cg_observable} for a qualitative understanding of the structural impact of the coarse-graining procedure, although the technical details may be skipped.
}
\subsection{The setup}\label{sec:observable_current_setup}
We define a set of observable (macroscopic) $\{o\}$ time-antisymmetric currents $\{J_o\}$ and the corresponding time-symmetric traffics $\{T_o\}$, with the many-to-one coarse-graining mapping $\mathcal{CG} : \{\gamma^{\rightleftharpoons} \} \to \{o\}$, thus $J_{o'} = \sum_{\{\gamma^\rightleftharpoons\}} \mathbb{O}_{o'\gamma} J_\gamma$ and $T_{o'} = \sum_{\{\gamma^\rightleftharpoons\}} \mathbb{O}_{o'\gamma} T_\gamma$, $\forall o' \in\{o\}$, where the matrix elements $\mathbb{O}_{o'\gamma} \in \{ 0, 1 \}$ of $\mathbb{O}$ defines the coarse-graining mapping $\mathcal{CG} :\{\gamma^{\rightleftharpoons} \} \to \{o\}$, which is represented mathematically as $\mathbb{O}_{o''\gamma'} \neq 0 \implies \mathbb{O}_{o' \gamma'} = 0, \forall o' \in \{o\} -o'', \forall \gamma' \in \{\gamma^\rightleftharpoons\}$. The support of observable $o'$ ($supp(o') = \{\gamma'| \mathbb{O}_{o'\gamma'} \neq 0\}$) defines the set of microscopic transitions, and its dimension quantifies the number of microscopic transitions $|N_{o'}| = |supp(o')|$ that contribute to $o'$. If observable currents account for all microscopic transitions, $|\{\gamma^\rightleftharpoons\}| = \sum_{o' \in \{o\}} |N_{o'}|$ holds. Introducing the vector notation $\vec{J}_o = \mathbb{O} \vec{J}_\gamma$ and $\vec{T}_o = \mathbb{O} \vec{T}_\gamma$. Physically, the binary-ness of $\mathbb{O}_{o'\gamma} \in \{ 0, 1 \}$ implies that a microscopic transition is either observable or not observable and imposes a scaling constraint on the observable EPR, ensuring that each microscopic transition is counted once in the observable currents. The many-to-one mapping constraint also ensures the linear independence of observable currents. 

\subsubsection{Main results}\label{sec:observable_current_results}
Using the contraction principle \cite{Touchette_2009}, we derive the observable Lagrangian/EPR $\mathcal{L}_{\{ o\}}^*[\{ J_{o}, T_{o} \}]/\dot{\Sigma}_{\{o\}}$ corresponding to observable currents and traffics in \cref{sec:contraction_principle_observable_current}, it reads,
\begin{equation}\label{eq:onsager_Machlup_functional_cg_observable}
    \dot{\Sigma}_{\{o\}} = \mathcal{L}_{ \{o\} }^*[\{ J_{o}, T_{o} \}]   
    = \sum_{ \{o\} } 2 J_o \tanh^{-1}{ \left( \frac{J_o}{T_o} \right) }.
\end{equation}
Using the bilinear form, $ \dot{\Sigma}_{\{o\}} = \sum_{ \{o\} } J_o \chi_o^*$, we obtain the inferred affinity $\chi_o^* = 2  \tanh^{-1}{ \left( {J_o}/{T_o} \right) }$ using $J_o$ and $T_o$. $ \mathcal{L}^*[\{ J_{\gamma}, T_{\gamma} \}]  \geq \mathcal{L}_{\{ o\}}^*[\{ J_{o}, T_{o} \}]$ holds due to the generalized log-normal inequality combined with the definitions of $J_o$ and $T_o$ \cite{Dannan_2014_log_sum_generalization}. Physically, it corresponds to the observable currents and traffics being able to capture a part of the microscopic EPR. Hence, in addition to using the exact rate functional, selecting all linearly independent microscopic observable currents is the second important criterion to obtain exact bounds on $\dot{\Sigma}$ using observable currents and traffics. Choosing $\{o\} = \{\gamma^\rightleftharpoons\}$ saturates the bound between $ \dot{\Sigma} $ and $ \dot{\Sigma}_{\{o\}} $.

Defining the scaled time-integrated stochastic observable current and the corresponding traffic, $\tilde{J}_{o} = \frac 1 \tau \int_0^\tau \hat{J}_o$ and $\tilde{T}_{o} = \frac 1 \tau \int_0^\tau \hat{T}_o$, respectively.
Integrating \cref{eq:onsager_Machlup_functional_cg_observable} from the initial $t=0$ to the final time $t=\tau$, the relation between the inferred EP $(\Sigma_o)$ and $\tilde{J}_{o}, \tilde{T}_{o}$ is,
\begin{equation}\label{eq:inferred_EP_and_thermodynamic_length}
    \tau \tilde{\Sigma}_{\{o\}} = \Sigma_{\{o\}} = \mathcal{S}_{\{o\}}^* = \int_0^{\tau} \mathcal{L}_{\{ o\}}^* d t \geq \sum_{ \{o\} } 2 \tau \tilde{J}_{o} \tanh^{-1}{ \left( \frac{ \tilde{J}_{o} }{ \tilde{T}_{o} } \right) }.
\end{equation}
\Cref{eq:onsager_Machlup_functional_cg_observable,eq:inferred_EP_and_thermodynamic_length} formulate the short- and finite-time non-quadratic thermodynamic lengths for the observable currents and traffics, analogous to \cref{eq:onsager_Machlup_functional,eq:EP_and_thermodynamic_length}, respectively. They relate the dynamical quantities: $\{J_o, T_o\}/ \{ \tilde{J}_o, \tilde{T}_o \}$ for short-time/finite-time to $(\dot{\Sigma}_{\{o\}})$/$({\Sigma}_{\{o\}})$. 

Using \cref{eq:inferred_EP_and_thermodynamic_length}, the transition probability measure for the observable current and traffic reads, 
\begin{equation}\label{eq:probability_observable_rate_functional}
\begin{split}
    \mathcal{P} \left[ \left\{ \tilde{J}_{o}, \tilde{T}_{o} \right\} \right] \asymp
    e^{-\tau I \left[ \left\{ { \tilde{J}_{o} }, { \tilde{T}_{o} } \right\} \right] },
\end{split}    
\end{equation}
\cref{eq:probability_observable_rate_functional} formulates the canonical ensemble using $ \{\tilde{J}_{o}, \tilde{T}_{o}\}$, analogously to \cref{eq:probability_exact_rate_functional}. Defining the precision of time-integrated observable current $\tilde{x}_o = \tilde{J}_o/\tilde{T}_o)$, the rate functional of the observable $(I[\tilde{J}_o, \tilde{T}_o])$ satisfies the scaling relation $I[\tilde{J}_o, \tilde{T}_o] = \tilde{T}_o f(\tilde{x}_o)$, implying that, for a fixed observation time $\tau$, the convergence of $\mathcal{P} \left[ \left\{ \tilde{J}_{o}, \tilde{T}_{o} \right\} \right]$ to the minimum of $(I[\tilde{J}_o, \tilde{T}_o])$ is accelerated due to larger values of $\tilde{T}_o$ compared to their microscopic counterparts. Thus, the relaxation time scale of $\tilde{T}_o$ is $\tau_o^{rel} = 1/\tilde{T}_o$, and is smaller than the microscopic counterparts. 

The derivation of variational formulations for observable currents relied on two key constraints. \textit{Constraint 1}: a binary notion of a microscopic current being observable or not observable, $\mathbb{O}_{o\gamma} \in \{0,1\}$. \textit{Constraint 2}: the many-to-one mapping ($\mathcal{CG} : \{\gamma^{\rightleftharpoons} \} \to \{o\}$) is used to obtain the coarse-grained currents. The validity of \cref{eq:onsager_Machlup_functional_cg_observable,eq:inferred_EP_and_thermodynamic_length,eq:probability_observable_rate_functional} holds even if these constraints are relaxed; see \cref{sec:general_observable_current}. 
\subsubsection{The condition for the saturation of the bound}
The condition for the saturation of equality between $ \dot{\Sigma}$ and $ \dot{\Sigma}_{\{o\}}$ is obtained using the log-normal inequality \cite{Dannan_2014_log_sum_generalization}. In particular, $\mathcal{L}^*[\{ J_{\gamma}, T_{\gamma} \}]  = \mathcal{L}_{\{ o\}}^*[\{ J_{o}, T_{o} \}]$, if $J_\gamma/T_\gamma = J_o/T_o$. Subsequently, this condition implies $\chi_\gamma^* = \chi_o^*, \forall \gamma \in supp(o)$. This can be easily verified using the bilinear form of the EPR, $\dot{\Sigma} = \sum_{\{ \gamma^{ \rightleftharpoons } \}} J_\gamma \chi_\gamma^* = \sum_{o} J_o \chi_o^* = \dot{\Sigma}_o$, if $\chi_\gamma^* = \chi_o^*, \forall \gamma \in supp(o)$. The bound $\dot{\Sigma} \geq \dot{\Sigma}_{o}$ is saturated if microscopic transition currents with equal effective affinities are counted together as a single observable current. Therefore, this bound is saturated for unicyclic graphs in steady state, where a single effective non-conservative affinity quantifies all non-conservative currents. However, for multi-cyclic systems, the affinities associated with all linearly independent cyclic currents need to be known, unless all of them are equal. 
{We briefly mention the relevance of this result in the context of Ref.\cite{Yang_2026_CFT}. A “Third Kirchhoff’s Law” for Stochastic Transport [equation (13) in Ref.\cite{Yang_2026_CFT}], which is a trivial sub-case of the condition $\chi_\gamma^* = \chi_o^*, \forall \gamma \in supp(o)$ obtained here for any generic multi-cyclic graph.
}
\subsection{Derivation}\label{sec:derivation_cg_observable} 
\subsubsection{Contraction principle approach}\label{sec:contraction_principle_observable_current}
In large deviation theory, the contraction principle deals with deriving the rate functional for an observable with knowledge of another rate functional \cite{Touchette_2009}. Here, we aim to apply the contraction principle to obtain the LDP for observable currents, given that the exact LDP for microscopic currents has been derived. The contraction of \cref{eq:onsager_Machlup_functional} under the constraint of the observable currents and traffics is formulated as the following constrained Lagrangian optimization problem:
\begin{equation}\label{eq:contraction_observable_lagrangian}
\begin{split}
    \mathcal{L}^*  \left[ \{ J_o, T_o \} \right] & = \inf_{ \{J_\gamma\}, \{T_\gamma\}, \{\lambda_{J_o}\}, \{\lambda_{T_o}\} } \bigg[ \mathcal{L}^*  \left[ \{ J_\gamma, T_\gamma \} \right] 
    \\
    & + \vec{\lambda}_{J_o} \cdot \left( \vec{J}_o - \mathbb{O} \vec{J}_\gamma \right) 
    + \vec{\lambda}_{T_o} \cdot \left( \vec{T}_o -  \mathbb{O} \vec{J}_\gamma \right)
    \bigg],
\end{split}    
\end{equation}
where, $\{\lambda_{J_o}\}$ and $\{\lambda_{T_o}\}$ are the Lagrange multipliers corresponding to the constraints imposed by $\{{J_o}\}$ and $\{{T_o}\}$, respectively. The extremization of \cref{eq:contraction_observable_lagrangian} with respect to $\lambda_{J_o}$ and $\lambda_{T_o}$ leads to the trivial constraint equations for $\{{J_o}\}$ and $\{{T_o}\}$. The optimization problem in \cref{eq:contraction_observable_lagrangian} for the set of observables $\{o\}$ is decoupled into independent optimization problems for  ${J_o}$ and ${T_o}$, due to the many-to-one coarse-graining mapping. 

Solving the decoupled optimization problem requires computing the Euler-Lagrange equations, $\forall \gamma \in \{\gamma^\rightleftharpoons\}$, given by:
\begin{subequations}\label{eq:euler_lagrange_microscopic}
\begin{equation} \label{eq:euler_lagrange_current}
    \frac{ \delta \mathcal{L}^*  \left[ \{ j_\gamma, T_\gamma \} \right] }{ \delta J_\gamma } 
    = 2\tanh^{-1} \left( \frac{J_\gamma}{T_\gamma} \right)
    + \frac{2 J_\gamma T_\gamma}{T_\gamma^2 - J_\gamma^2} - \lambda_{J_{o}} \mathbb{O}_{o\gamma}.  
\end{equation}
\begin{equation} \label{eq:euler_lagrange_traffic}
    \frac{ \delta \mathcal{L}^*  \left[ \{ j_\gamma, T_\gamma \} \right] }{ \delta T_\gamma } 
    = - \frac{2 J_\gamma^2 }{T_\gamma^2 - J_\gamma^2} - \lambda_{T_{o}} \mathbb{O}_{o\gamma}.  
\end{equation}
\end{subequations}    
where, \cref{eq:euler_lagrange_microscopic} holds $\forall \gamma \in \{\gamma^\rightleftharpoons\}$, and maps $\gamma$ to a unique $o \in \{o\}$. Solving the optimization problem trivially implies ${ \delta \mathcal{L}^*  \left[ \{ j_\gamma, T_\gamma \} \right] }/{ \delta J_\gamma } = 0$ and ${ \delta \mathcal{L}^*  \left[ \{ j_\gamma, T_\gamma \} \right] }/{ \delta T_\gamma } = 0$,
\begin{subequations}\label{eq:euler_lagrange_microscopic_final}
\begin{equation} \label{eq:euler_lagrange_current_final}
    2\tanh^{-1} \left( \frac{J_\gamma}{T_\gamma} \right)
    + \frac{2 J_\gamma T_\gamma}{T_\gamma^2 - J_\gamma^2} = \lambda_{J_{o}} \mathbb{O}_{o\gamma}.  
\end{equation}
\begin{equation} \label{eq:euler_lagrange_traffic_final}     
 - \frac{2 J_\gamma^2 }{T_\gamma^2 - J_\gamma^2} = \lambda_{T_{o}} \mathbb{O}_{o\gamma}.  
\end{equation}
\end{subequations}    
Computing $J_\gamma \times$\cref{eq:euler_lagrange_current_final} + $T_\gamma \times$ \cref{eq:euler_lagrange_traffic_final} leads to $2 J_\gamma \tanh^{-1} \left( {J_\gamma}/{T_\gamma} \right) 
= \mathbb{O}_{o\gamma} \left( \lambda_{J_{o}} J_\gamma  + \lambda_{T_o} T_\gamma \right) $. This simplifies $\inf_{ \{J_\gamma, T_\gamma\} } \mathcal{L}^*  \left[ \{ J_\gamma, T_\gamma \} \right] = \sum_{\{\gamma^\rightleftharpoons\}} O_\gamma \left( J_\gamma \lambda_{J_o} + T_\gamma \lambda_{T_o} \right)$ to $ \mathcal{L}^*  \left[ \{ J_o, T_o \} \right] = \sum_{\{o\}} \left(  J_{o} \lambda_{J_o} +  T_{o} \lambda_{T_o} \right) = \vec{\lambda}_{J_o}^T \vec{J}_o + \vec{\lambda}_{T_o}^T \vec{T}_o$ in its bilinear form. 

For $\gamma, \gamma' \in supp(o)$, the right-hand side of \cref{eq:euler_lagrange_current_final} or \cref{eq:euler_lagrange_traffic_final} is equal to $\lambda_{J_o}$. Hence, it imposes equality on the left-hand side of \cref{eq:euler_lagrange_current_final,eq:euler_lagrange_traffic_final} for $\gamma, \gamma' \in supp(o)$. However, the left-hand sides are monotonic functions of $x_\gamma = J_\gamma / T_\gamma$, defined as $b_1(x) = 2\tanh^{-1}(x) + 2x/(1-x^2)$ and $b_2(x) = 2x^2/(1-x^2)$. This implies $\forall \gamma, \gamma' \in supp(o)$, the solution of \cref{eq:contraction_observable_lagrangian} satisfies $ x_\gamma = x_{\gamma'} = x_{o} = J_{o}/T_{o}$. Plugging it into \cref{eq:contraction_observable_lagrangian} leads to 
\begin{equation}
    \mathcal{L}^*  \left[ J_o, T_o \right] = 2 J_o \tanh^{-1} \left(\frac{J_{o}}{T_{o}} \right), \hspace{1cm} \forall o \in \{o\},
\end{equation}
hence, $\mathcal{L}^*  \left[ \{ J_o, T_o \} \right] = \sum_{\{o\}} 2 J_o \tanh^{-1}(J_{o}/T_{o})$. Therefore, $\chi_o^* = 2 \tanh^{-1}(J_{o}/T_{o})$ corresponds to the effective transition affinity of the observable current, which can be equivalently inferred using $J_o$ and $T_o$. Comparing $\chi_o^*$ to the bilinear form derived previously $ \mathcal{L}^*  \left[ \{ J_o, T_o \} \right] = \vec{\lambda}_{J_o}^T \vec{J}_o + \vec{\lambda}_{T_o}^T \vec{T}_o$, it is equal to an effective Lagrange multiplier for $J_o$ with $\chi_o^* = \lambda_{J_o} + \lambda_{T_o} T_o/J_o$. 
\subsubsection{Linear algebra approach}\label{sec:general_observable_current}
For partial (scaled) observation of transition currents, such that $\mathbb{O}_{o\gamma} \in \mathbb{R}^+$ is any real number and not necessarily $\mathbb{O}_{o\gamma} \in \{0,1\}$, we note that to respect the scale invariance of the EPR $\dot{\Sigma}$, $\mathcal{L}[\{J_\gamma, T_\gamma \}]$ has to be invariant under the scaling transformation of currents and traffics $J_\gamma \to \bar{J}_\gamma = c J_{\gamma}$ and $T_\gamma \to  \bar{T}_\gamma = c {T}_\gamma$ \footnote{We use an overbar to denote a scaling transformation, avoiding confusion with scaled time-integrated physical quantities denoted using a tilde.}, implying scaling transformation $x_\gamma \to \bar{x}_\gamma = x_\gamma$. Hence, our naive calculation implies that the effective affinity is also invariant under the scaling transformation, $\chi_\gamma^* \to \bar{\chi}_\gamma^* = \chi_\gamma^*$. Using the bilinear form of the EPR, $\dot{\Sigma} = \sum_{\gamma^\rightleftharpoons} J_\gamma \chi_\gamma^*$, we reach the scaling transformation of the EPR, $\dot{\Sigma} \to  \dot{\bar{\Sigma}} = c \dot{\Sigma}$. However, this contradicts the invariance of the EPR under the scaling transformation of currents. 

To address this problem associated with scaling, one notices that a simultaneous scaling of time, $t \to \bar{t} = t/k$, restores the scaling invariance of the EPR, also seen equivalently by the scaling of traffic $T_\gamma \to \bar{T}_\gamma = k T_\gamma$ that defines the inverse timescale for $\gamma^\rightleftharpoons$. This key physical insight implies that \cref{eq:precision_bound} is invariant under scaling transformation and holds for scaled microscopic currents, therefore, the proof derived in \cref{sec:contraction_principle_observable_current} is extended by relaxing \textit{Constraint 1}. This amounts to applying the contraction principle to $\bar{J}_\gamma = \mathbb{O}_{o\gamma} J_\gamma$. This symmetry has previously been realized on the Lagrange multipliers $\lambda_{J_o}$ and $\lambda_{T_o}$ in the right-hand side of \cref{eq:euler_lagrange_microscopic_final}, where $\chi_{o}^*$ depends linearly on $\lambda_{J_o}$ and $\lambda_{T_o}$ with a constant multiplied by $\mathbb{O}_{o\gamma}$.

Similarly, we exploit the scale invariance of EPR to extend the variational formulation for observable currents by relaxing \textit{Constraint 2}. For this purpose, the linear algebraic formulation implies, to the preserve the invariance of $\dot{\Sigma}_{\{o\}}$ $J_\gamma \to \bar{J}_\gamma = k J_\gamma$ should be compensated by $\chi_\gamma^* \to \bar{\chi}_\gamma^* = \chi_\gamma^*/k$ defined for microscopic currents. To this end, we utilize the bilinear form of the EPR, $\dot{\Sigma} = \sum_{\{ \gamma^\rightleftharpoons\}} J_\gamma \chi_\gamma^* = (\vec{\chi}_\gamma^*)^T \vec{J}_\gamma$. Thus, the problem is reduced to a norm-preserving basis transformation in linear algebra. How do effective observable affinities transform under the dual transformation that preserves the correct scaling of $\dot{\Sigma}$ using $\dot{\Sigma}_{\{o\}}$?, given that microscopic currents transform as: $\vec{J}_o = \mathbb{O} \vec{J}_\gamma$ with $\mathbb{O}_{o'\gamma} \in  \mathbb{R}^+$. By not assuming any constraint on the structure of the matrix $\mathbb{O}$, the linear algebraic formulation of the problem inherently violates \textit{Constraint 2}. Therefore, the effective affinity of the observable current should transform as $(\vec{\chi}_o^*)^T = (\vec{\chi}_\gamma^*)^T \mathbb{O}^\dagger$, where $\mathbb{O}^{\dagger} = \mathbb{O}^T ( \mathbb{O} \mathbb{O}^T )^{-1}$ is the right pseudo-inverse that dictates the transformation of $(\vec{\chi}_\gamma^*)^T$ in the dual conjugate space to the current space. This leads to $\dot{\Sigma}_{\{o\}} = \mathcal{L}_{ \{o\} }^*[\{ J_{o}, T_{o} \}] = \sum_{\{o\}} J_o \chi_o^* = (\vec{\chi}_o^*)^T \vec{J}_o$.

Using the linear algebraic solution, we highlight the consequences of the two constraints on the structure of effective observable affinities. 
Due to \textit{Constraint 2}: the observable conjugate affinities simplify to $\chi_{o}^* = \sum_{\gamma \in supp(o) } \chi_{\gamma}^*\mathbb{O}_{o\gamma} /||{O}_{o}|| $, where $||O_o|| = \sum_{\gamma} \mathbb{O}_{o\gamma}^2$, since $\mathbb{O} \mathbb{O}^T$ is decomposed into a block-diagonal form, resulting in decoupling of $\chi_o$ for linearly independent observable currents. Due to \textit{Constraint 1}: $\chi_{o}^* = \sum_{\gamma^\rightleftharpoons \in supp(o) } \chi_\gamma^* /|N_o|$. Violation of \textit{Constraint 2}: in this case, $\{J_o\}$ are not linearly independent due to $\mathbb{O}_{o^1\gamma'} \neq 0 \centernot \implies \mathbb{O}_{o^2\gamma'} = 0$. Although $\chi_o^*$ is obtained using $\mathbb{O}^\dagger$, a simplified closed-form expression for $\chi_o^*$ is not available compared to previous cases. Because, due to cross-coupling terms, a microscopic transition current contributes to multiple observable currents, which leads to non-vanishing non-diagonal terms in $(\mathbb{O}\mathbb{O}^T)^{-1}$. This cross-coupling is transferred from the transformation $\{\chi_\gamma^*\} \to \{\chi_o^*\}$, making it rather difficult to visualize the transformation. However, a system-specific brute-force or numerical computation of $\mathbb{O}^\dagger$ is always feasible.
\section{Manifestations of Minimum Action Principle}\label{sec:applications_min_ac_princpl}
\subsection{Single observable currents: stochastic EPR is the most precise current}\label{sec:fr_single_observable_current}
For a single observable current, $\mathbb{O}$ is reduced a vector $\vec{O}$, which leads to $( \vec{O} \vec{O}^{T} ) = \sum_{ \{\gamma^\rightleftharpoons\} } (O_\gamma)^2$, therefore, $\chi_o^* = \left[ \sum_{ \{\gamma^\rightleftharpoons\} } \chi_\gamma^* O_\gamma \right] / \left[ \sum_{ \{\gamma^\rightleftharpoons\} } (O_\gamma)^2 \right]$. Choosing $\vec{O} = \vec{\chi}_\gamma^*$ corresponds to the stochastic EPR as an observable current; which leads to $\chi_o^* = 1$. Imposing the normalization of the observable current, $\sum_{ \{\gamma^\rightleftharpoons\} } (O_\gamma)^2 = 1$, we investigate the condition to observe the most precise observable current. This amounts to solving the variational optimization problem: $\chi_o^{max} = \sup_{\vec{O}} (\chi_o)$ under the normalization constraint. Therefore, $\delta \chi_o^*/\delta O_\gamma = 0, \forall \gamma \in \{ \gamma^{\rightleftharpoons}\}$, which implies $O_{\gamma}/\chi_\gamma^* = O_{\gamma'}/\chi_{\gamma'}^* = \text{constant}, \forall \gamma, \gamma' \in \{ \gamma^{\rightleftharpoons}\}$. Using the normalization constraint, its unique optimal solution is $O_{\gamma} = \chi_{\gamma}^* / ||\chi^*||$ or $\vec{O}_{\gamma} = \vec{\chi}_{\gamma}^* / ||\chi^*||$ and $\chi_o^{max} = ||\chi^*||$, where $||\chi^*|| = \sqrt{\sum_{ \{\gamma^\rightleftharpoons\} } (\chi_\gamma^*)^2}$ denotes the absolute value or the length.

Physically, this implies that, among all normalized single observable currents $J_o$, the maximum observable affinity corresponds to choosing a current along (parallel to) the stochastic EPR, $J_o = \dot{\Sigma}_{st}/||\chi^*||$ with an effective affinity $\chi_{\tilde{\Sigma}_{st}} = \chi_o^{max}/||\chi^*|| = 1$. Analogously formulated, for the thermodynamic inference using a single observable current, the stochastic EPR is the most precise observable current that maximizes the quantification of non-equilibrium-ness, it is a `special' current that exactly quantifies the microscopic thermodynamic dissipation on the macroscale, which is otherwise lost due to the suboptimal choice $J_o$. The least optimal observable current for thermodynamic inference is orthogonal to the stochastic EPR. Since $J_o \perp \vec{\chi}_\gamma^*$, in this case, $\chi_o^* = 0$, because $\sum_{\{\gamma^\rightleftharpoons\}} O_\gamma \chi_{\gamma}^* = 0$. These results hold independently of the dimension of $\{\gamma^\rightleftharpoons\}$ and the steady-state assumption for any observation time $\tau$, and apply to any multi-cyclic system. 
\subsection{Partial control description}
If the transition affinities are known, then $\chi_\gamma^* = A_\gamma$, this corresponds to the partial control description of MinAP. Then $\mathcal{L}_{\{o\}}^*$ is reduced to its bilinear form. Here, the partial control description refers to the control of the transition affinities, which are fixed and known, but the transition currents are still the `uncontrollable' stochastic observables.
\subsubsection{Fluctuation relation}\label{sec:fluctuation_relation}
One observes that the normalization condition for the probability distribution \cref{eq:probability_observable_rate_functional} trivially implies the integrated FR $\langle e^{-\tau (\vec{\chi}_o^*)^T \vec{\tilde{J}}_o} \rangle = 1$. Defining the Scaled Cumulant Generating Function (SCGF) for observable currents, $\mathcal{K}_{\vec{\tilde{J}}_{o}}(\chi_o) = \lim_{\tau \to \infty} \frac{1}{\tau} \ln{ \langle e^{\tau \vec{\chi}_o \cdot \vec{\tilde{J}}_o} \rangle }$ \cite{Touchette_2009}, where, SCGF is defined with respect to the large deviation parameter. SCGF defines the non-equilibrium analog of the system's free energy. Using $\mathcal{P}[\vec{J}_o] \asymp e^{-\tau \vec{\chi}_o^* \cdot \vec{\tilde{J}}_o}$,  it is known SCGF satisfies the Gallavotti-Cohen FR symmetry: $\mathcal{K}_{\vec{\tilde{J}}_{o}} \left( \vec{\chi}_o \right) = \mathcal{K}_{ \vec{\tilde{J}}_{\alpha}} \left( - \vec{\chi}_o^* - \vec{\chi}_o \right)$ \cite{Gallavotti_1995,Touchette_2009,Andrieux_2007,Andrieux_2007_single_current_FT} and reveals the asymmetry/symmetry of the detailed FR. Since $\chi_o^*$ corresponds to the non-trivial root of SCGF, thereby quantifying the detailed FR symmetry for the time-integrated observable stochastic currents $\vec{\tilde{J}}_o$, \cite{Andrieux_2007,Andrieux_2007_single_current_FT,Touchette_2009}.
\begin{equation}\label{eq:detailed_FR_observable_curent}
    \ln \left( \frac{ \mathcal{P} [ \vec{\tilde{J}}_o = \langle \vec{\tilde{J}}_o \rangle ] }{ \mathcal{P} [ \vec{\tilde{J}}_o = -\langle \vec{\tilde{J}}_o \rangle ] } \right) = \tau \vec{\chi}_o^* \cdot \langle \vec{\tilde{J}}_o \rangle = \mathcal{S}_{\{o\}}^*,
\end{equation}
where, $\mathcal{P}[\vec{\tilde{J}}_o = \langle \vec{\tilde{J}}_o \rangle ]$ is the shorthand notation for the probability density of observing a value $\langle \vec{J}_o \rangle$ for $\vec{\tilde{J}}_o$.  

We consider the notable case of observable currents. First, the stochastic EPR as an observable current, $J_o = \dot{\Sigma}_{st} = \sum_{\{\gamma^\rightleftharpoons\}} \chi_\gamma^*$, which results in $\chi_o^* = 1$, derived in \cref{sec:fr_single_observable_current}, leading to the Lebowitz-Spohn symmetry: $\mathcal{K}_{\tilde{\Sigma}_{st}}\left( \chi_{\tilde{\Sigma}_{st}} \right) = \mathcal{K}_{\tilde{\Sigma}_{st}} \left( - 1 - \chi_{\tilde{\Sigma}_{st}} \right)$, and the detailed FR for $\dot{\Sigma}_{st}$,
\begin{equation}\label{eq:detailed_FR_stochastic_EP}
    \ln \left( \frac{ \mathcal{P} [ {\tilde{\Sigma}}_{st} = \langle {\tilde{\Sigma}}_{st} \rangle ] }{ \mathcal{P} [ {\tilde{\Sigma}}_{st} = -\langle {\tilde{\Sigma}}_{st} \rangle ] } \right) = \tau \langle {\tilde{\Sigma}}_{st} \rangle, 
\end{equation}
and the integrated FR for $\dot{\Sigma}_{st}$ is $\langle e^{ -\tau \tilde{\Sigma}_{st} } \rangle = 1$ \cite{Lebowitz_1999,Touchette_2009}. Second, if we consider the most fundamental case of all microscopic transition currents, $\{o\} = \{\gamma^\rightleftharpoons\}$, with known transition affinities, $\{\chi_\gamma^*\} = \{A_\gamma\}$. The detailed FR for $\{J_\gamma\}$ is,
\begin{equation}\label{eq:detailed_FR_microscopic_curents}
\begin{split}
    \ln \left( \frac{ \mathcal{P} [ \vec{\tilde{J}}_\gamma = \langle \vec{\tilde{J}}_\gamma \rangle ] }{ \mathcal{P} [ \vec{\tilde{J}}_\gamma = -\langle \vec{\tilde{J}}_\gamma \rangle ] } \right) = \tau \vec{A}_\gamma \cdot \langle \vec{\tilde{J}}_\gamma \rangle.
\end{split}    
\end{equation}
and the integrated FR for $J_\gamma$ is $\langle e^ {- \tau A_\gamma \tilde{J}_\gamma} \rangle = 1$.
\subsubsection{The effective affinity and martingale property}
 Effective affinity $\chi_\gamma^*$ plays a key role in quantifying FR symmetry. Here, we highlight its underlying mathematical structure, namely, the martingale property \cite{Roldan_2023,Chetrite_2011,Neri_2017,Pigolotti_2017,Chetrite_2019,Chun_2019,Neri_2022,Roldan_2023,Raghu_2025,Neri_2025}, which has important implications for thermodynamic inference in the absence of observable currents \cite{Martinez_2019,Skinner_2021,Skinner_2021_pnas}, using first-passage time statistics or waiting-time statistics \cite{Neri_2017,Neri_2025,Raghu_2025}, and applications to thermodynamic inference \cite{Nadal_rosa_2025}. 

TL satisfies the additive property $ \tau \tilde{J}_o[\tau] = \tau' \tilde{J}_o[\tau'] + (\tau - \tau') \tilde{J}_o [\tau - \tau']$, with the initial value condition $\tilde{J}_o[\tau] = 0$. If $\chi_o^*$ is time-homogeneous, that is, $\chi_o^*[\tau] = \chi_o^*[\tau'] = \chi_o^*[\tau-\tau']$, then the action satisfies the additive property $\mathcal{S}_{\{o\}}^*[\tau] = \mathcal{S}_{\{o\}}^*[\tau'] + \mathcal{S}_{\{o\}}^*[\tau - \tau']$, with the initial value condition $ \mathcal{S}_{\{o\}}^*[\tau] = 0$. The integrated FR $\langle e^{-\mathcal{S}_{\{o\}}^*} \rangle = 1$ is also satisfied. These are sufficient conditions for $e^{-\mathcal{S}_{\{o\}}^*}$ ( or equivalently $e^{-\Sigma_{\{o\}}}$), to be a martingale \cite{Roldan_2023}. 
{The time-homogeneity of $\chi_o^*$ required for the martingale property of $J_o$ is equivalent to the observable being memory-less; a regime in which the LDB condition effectively emerges from MinAP as discussed in \cref{sec:beyond_LDB}. This results in the linear scaling of $\mathcal{S}_{\{o\}}^*[\tau]$ in $\tau$, ensuring the time-additive property, and violations of it due to transient effects need to be studied \cite{Evans_2004,Evans_2005}.} 
From the most fundamental perspective, if we choose all microscopic currents $\{o\} = \{ \gamma^\rightleftharpoons \}$, the exponentiated negative of $J_\gamma$ is a Martingale with effective affinity $\chi_\gamma^*$ that characterizes the directional asymmetry between observation of the positive and negative amplitudes of its value. This physical property has been realized earlier through the detailed FR symmetry \cref{eq:detailed_FR_microscopic_curents} for microscopic currents. Therefore, the martingale property of the microscopic transition currents is the most fundamental thermodynamic symmetry, the stochastic EP is a `special' case, whose martingale property has been rigorously studied \cite{Roldan_2023,Chetrite_2011,Neri_2017,Pigolotti_2017,Chetrite_2019,Chun_2019}.  

In the measure-theoretic formalism, the Radon-Nikodym derivative (RND) is defined here as the ratio of the transition probability measure between the forward and backwards processes, which by definition is the exponential of left side of \cref{eq:detailed_FR_observable_curent,eq:detailed_FR_stochastic_EP,eq:detailed_FR_microscopic_curents}. The equivalence between the stochastic EP (a physical property) and the logarithm of RND (a mathematical property) assigns a thermodynamic meaning to it \cite{Yang_2020,Hong_2020_measure_theory,Qian_2019,Jiang_2003}. It unveils the FR symmetry and martingale property of stochastic EP \cite{Yang_2020,Hong_2020_measure_theory,Qian_2019,Jiang_2003}. Here, we extend the measure-theoretical formalism to the most fundamental microscopic transition currents of discrete-state processes \cref{eq:detailed_FR_microscopic_curents}, so that the existing measure-theoretical understanding of stochastic EP is recovered by the contraction of microscopic currents to stochastic EP (\cref{eq:detailed_FR_microscopic_curents} to \cref{eq:detailed_FR_stochastic_EP}) \cite{Yang_2020,Hong_2020_measure_theory,Qian_2019,Jiang_2003}. We have briefly outlined the connection to martingale structure and measure-theoretical formulation; however, a more systematic and rigorous mathematical analysis, as well as stronger implications of martingale theory \cite{Roldan_2023} or measure theory  \cite{Yang_2020} for other physical observables and practical applications such as thermodynamic inference \cite{Nadal_rosa_2025} remain to be explored. 
\subsubsection{Orthogonal decomposition of EPR}\label{sec:orthogonal_epr}
Time-integrated relaxation and dissipative currents satisfy the scalings $\int_0^\tau dt {J}_\gamma^{rel} \propto O(1)$ and $\int_0^\tau dt {J}_\gamma^{ss} \propto \tau$, respectively. Physically, this results in short-time and long-time symmetries of currents for the relaxation and dissipative currents, respectively. The origin of the orthogonality of relaxation and steady state is attributed to the decomposition of $A_\gamma = A_\gamma^{rel} + A_\gamma^{ss}$ into its boundary and bulk terms, respectively, which results in the excess and housekeeping EPR scalings, $\Sigma^{ex} \propto O(1)$ and $\Sigma^{hk} \propto O(\tau)$. However, we have assumed a dissipative scaling of $\{J_\gamma\}$ to obtain the LDP from the variational formulation. Thus, the LDP must be modified to accurately account for the different scalings of the relaxation and dissipative currents, thereby restoring the short-time symmetry of the stochastic currents.

Consider the orthogonal decomposition of the transition affinities $\chi_\gamma^* = \chi^{ss}_\gamma + \chi^{ex}_\gamma$, such that, $\chi^{ss}_\gamma = F_\gamma$ and $\chi_\gamma^{ex} = \Delta_\gamma \chi_{i}^{ex}$ with $\chi^{ex}_i = -\ln{(\rho_i/\rho_i^{E})} = S_i^{state|E}$, where $\chi_\gamma^{ex}$ characterizes the  distance from the Boltzmann distribution $\rho_i^{E} = e^{-E_i + \psi_E}$. Using the linearly independent decomposition of the transition affinities into relaxation and dissipative components, we define, $\vec{J}_{o} = ( - \partial_t \psi_E, -\sum_{ \{i\} } \ln(\rho_i/\rho_i^E), \vec{F}_\gamma )$, with the respective scaling vector $\vec{\Omega} = ( 1, 1, \vec{\tau} )$. Furthermore, we consider the total dissipative transition affinity as a single dissipative current. Therefore, $\vec{J}_{o} = ( - \partial_t \psi_E, -\sum_{ \{i\} } \ln(\rho_i/\rho_i^E), \sum_{\{\gamma^\rightleftharpoons\}} {F}_\gamma )$, with the scaling vector $\vec{\Omega} = ( 1, 1, \tau )$. The third term here implies, using the observable transition current $F_\gamma$, when a transition $\gamma^\rightleftharpoons$ takes place. Hence, the choice of $\vec{J}_{o}$ is equivalent to the orthogonal decomposition of the stochastic EPR $\vec{\dot{\Sigma}}_{st} = ( \dot{W}_{qs}, \sum_{i} \hat{S}_i^{state|E}, \dot{\Sigma}^{hk})$ with  $\vec{\chi}_{st}^* = ( 1, 1, 1)$. Here, the stochastic EPR $\vec{\dot{\Sigma}}_{st}$ is decomposed into three linearly independent contributions. First, the quasistatic driving work rate $\dot{W}_{qs}$, a boundary term in the control parameter space $\{\lambda\}$ of $E(\{\lambda\})$, which depends on the explicit time-dependent driving of $\{\lambda\}$. Second, the relaxation/excess EPR, a boundary term in the probability state-space $\{\rho_i\}$, which is the statistical distance between the initial and final states relative to the reference Boltzmann distribution. Third, the housekeeping EPR, a bulk term that scales with $\tau$ and is supported by a dissipative bath: a non-vanishing dissipative EP contribution in the steady state. Thus, a detailed FR symmetry for the orthogonal decomposition of the EP reads
\begin{equation}\label{eq:detailed_fluctuation_relation}
\begin{split}
    \ln \left( \frac{\mathcal{P}[ \vec{\tilde{\Sigma}}_{st} 
    = \langle \vec{\tilde{\Sigma}}_{st} \rangle ]}
    {\mathcal{P}[ \vec{ \tilde{\Sigma}}_{st} = -\langle \vec{\tilde{\Sigma}}_{st} \rangle ]} \right) 
    = \vec{\chi}_{st}^* \cdot \vec{\Omega} \odot \langle \vec{\tilde{\Sigma}}_{st} \rangle,
\end{split}    
\end{equation}
where, $\odot$ denotes a component-wise Hadamard product defined between $\vec{\Omega}$ and $\vec{\tilde{\Sigma}}_{st}$, simplified to obtain the total time-integrated EP, $\chi_{st}^* \cdot \vec{\Omega} \odot \langle \vec{\tilde{\Sigma}}_{st} \rangle = \langle -\Delta_0^\tau \psi_E + \Delta_0^\tau S_{state}^E +  \tau \tilde{\Sigma}_{hk} \rangle$.
\subsection{Thermodynamic inference description}\label{sec:thermodynamic_inference}
We show applications of the MinAP to thermodynamic inference, and discuss three cases using state-space observables, which are experimentally easily assessable \cite{Terlizzi_2024}, compared to TKUR which requires current statistics.
\subsubsection{Non-quadratic speed limit}\label{sec:speed_limits}
If observable currents $\{o\} = \{J_i\}$ into the state $\{\rho_i\}$ are chosen. This choice of observable currents corresponds to the contraction from the transition-space to the state-space through the continuity equation, $\partial_t \rho_i = J_i$. The traffic defined for the state $\rho_i$ is $T_i = \sum_{i \in \{\gamma^\rightleftharpoons\} } T_{\gamma}$ and quantifies the total scaled variance of $\rho_i$ due to microscopic transitions (where $\rho_i$ is involved). The time-integrated continuity equation implies $\Delta_0^\tau \rho_i = \rho_i(\tau) - \rho_i(0) = \tau \tilde{J}_i$.
This reduces \cref{eq:inferred_EP_and_thermodynamic_length} to
\begin{equation}\label{eq:non_quadratic_speed_limit}
    \Sigma_{SL} = 2 \Delta_0^\tau \rho_i \tanh^{-1}{ \left( \frac{ \Delta_0^\tau \rho_i }{ \tau \tilde{T}_{i} } \right) },
\end{equation}
\cref{eq:non_quadratic_speed_limit} is a non-quadratic speed limit and generalizes the quadratic speed limit from Ref.\cite{Yoshimura_2021}. Here, $\tilde{T}_i$ is the scaled diffusion constant for $\rho_i$. The approximation $\tanh^{-1}(x) \approx x$ gives the quadratic speed limit in its more familiar form \cite{Yoshimura_2021}.  However, the mismatch increases for fEQ systems, as discussed earlier.
\subsubsection{\label{sec:onsager_machlup_functional} Non-quadratic Onsager-Machlup functional and fluctuations around steady state}
We aim to use the MinAP to study the fluctuations around the steady state. For this purpose, we choose the relaxation currents of states $\{i\}$, $\{J_o\} = \{J_i^{rel}\}$. Using the orthogonal decomposition $\{J_i\}$ into dissipative and relaxation currents, the continuity equation $\partial_t \rho_i = J_i^{rel} + J_i^{ss}$, and the short-time non-quadratic TKUR \cref{eq:onsager_Machlup_functional_cg_observable}, we obtain the excess Lagrangian $\mathcal{L}_{rel}^*$ for fluctuations around the steady state,
\begin{equation}\label{eq:relaxation_lagrangian}
    \mathcal{L}_{rel}^* = \sum_{\{i\}} 2 (\partial_t \rho_i - J_i^{ss}) \tanh^{-1}{ \left( \frac{\partial_t \rho_i - J_i^{ss}}{\tau \tilde{T}_i} \right) }.
\end{equation}
with a non-quadratic Onsager-Machlup functional for the probability distribution of the fluctuations around the steady state,
\begin{equation}\label{eq:OM_relaxation_probability_functional}
\begin{split}
    \mathcal{P}[\{\rho_i, J_i^{ss}\}] \asymp e^{- \int_0^\tau \mathcal{L}_{rel}^*  dt }.
\end{split}    
\end{equation}
The Gaussian approximation $\tanh^{-1}{(x)} \approx x$ of \cref{eq:relaxation_lagrangian,eq:OM_relaxation_probability_functional} leads to the quadratic Onsager-Machlup functional \cite{Onsager_1953,Onsager_1953_2}, derived originally for Gaussian fluctuations around the equilibrium steady state but generalized here with \cref{eq:relaxation_lagrangian,eq:OM_relaxation_probability_functional} for any non-equilibrium steady state $\{J_i^{ss}\}$ and incorporating non-Gaussian fluctuations. 

Importantly, if the relaxation-fluctuation symmetry is satisfied, $\mathcal{L}_{rel}^* = -d_t D_{ss}^{KL}$. However, this is not generally the case, since \cref{eq:relaxation_lagrangian} assigns a non-quadratic thermodynamic EPR cost to fluctuations around the steady state. In contrast, $-d_t D_{ss}^{KL}$ is the thermodynamic EPR cost associated with the relaxation process (the gradient descent) towards the steady state, with $-D_{ss}^{KL}$ being the corresponding Lyapunov functional for the relaxation process. Since the relaxation-fluctuation symmetry is not necessarily satisfied in fEQ systems, we clarify these differences between the fluctuations around the steady state and the relaxation towards the steady state, governed by $\mathcal{L}_{rel}^*$ in \cref{eq:relaxation_lagrangian} and $-d_t D_{ss}^{KL}$, respectively. 
\subsubsection{Non-quadratic state-space TKUR}\label{sec:state_space_tkur}
A novel class of systems that breaks the `actio=reactio' symmetry are called `non-reciprocal systems' and manifests the formation of vorticity currents, defined between two states is defined as $\omega_{ij} = \rho_j \partial_t \rho_i - \rho_i \partial_t \rho_j$ \cite{ATM_2024_nr_st}. Importantly, $\omega_{ij}$ is analog of an non-equilibrium current, which is defined in state-space and does not require knowledge of the underlying topology of transitions (graph). This makes thermodynamic inference using $\omega_{ij}$ suitable and appealing for experimental purposes  \cite{Terlizzi_2024,Tomita_1974,Terlizzi_2024_long,Ohga_2023,Shiraishi_2023}.  

We choose $\rho_j J_i$ as the unidirectional observable current, which leads to a bidirectional current $J_o = \rho_i J_j - \rho_j J_i$. Using the continuity equation, $\omega_{ij} = \rho_i \partial_t \rho_j - \rho_j \partial_t \rho_i = J_o$. The corresponding observable traffic is defined as $\omega_{ij}^s = T_o = \rho_i \partial_t \rho_j + \rho_j \partial_t \rho_i$. Choosing the set of all possible combinations of state pairs $n(n-1)/2$ leads to all linearly independent vorticity currents as observable currents in state-space, $\{J_o\} = \{\omega_{ij}\}$. Defining the temporal state correlations, $C_{ij}(\tau) = \rho_i(\tau) \rho_j(0)$, between $\rho_i$ and $\rho_j$ over time $\tau$ \cite{Tomita_1974} and decomposing into its state-symmetric and state-anti-symmetric components, $C_{ij}^s(\tau) = \rho_i(\tau) \rho_j(0) + \rho_j(\tau) \rho_i(0)$ and $C_{ij}^a(\tau) = \rho_j(\tau) \rho_i(0) - \rho_i(\tau) \rho_j(0)$, respectively. The time-integrated vorticity $\tau \tilde{\omega}_{ij} = \int_0^\tau \omega_{ij} dt = \Delta_0^\tau C_{ij}^a(\tau)$ and time-integrated traffic $\tau \tilde{\omega}_{ij}^s = \int_0^\tau \omega_{ij}^s dt = \Delta_0^\tau C_{ij}^s(\tau)$ are simplified, where $\Delta_0^\tau C_{ij}^a(\tau) = C_{ij}^a(\tau) - C_{ij}^a(0)$ and $\Delta_0^\tau C_{ij}^s(\tau) = C_{ij}^s(\tau) - C_{ij}^s(0)$ quantify the change over observation time $\tau$. 

Hence, using \cref{eq:inferred_EP_and_thermodynamic_length} for $\{J_{o}\} = \{\omega_{ij}\}$ leads to,
\begin{equation}\label{eq:inferred_EP_and_thermodynamic_length_correlation}
\begin{split}   
    {\Sigma}_{\{\omega\}} = \int_0^{\tau} \mathcal{L}_{\{ \omega \}}^* d t & \geq \sum_{ \{ij\} } 2 \Delta_0^\tau {C}_{ij}^a \tanh^{-1}{ \left( \frac{ \Delta_0^\tau {C}_{ij}^a }{ \Delta_0^\tau {C}_{ij}^s } \right) }.
\end{split}
\end{equation}
\Cref{eq:inferred_EP_and_thermodynamic_length_correlation} is the non-quadratic state-space TKUR quoted in Ref.\cite{ATM_2024_nr_st} for non-reciprocal systems. It obtains a bound on $\Sigma$ using using state-space temporal correlations ${C}_{ij}^a(\tau)$ and ${C}_{ij}^s(\tau)$ (instead of the usual current-space formulation). The choice $\{J_o\} =\{\omega_{ij}\}, \forall i, j \in {\{i\}} $ obtains the tightest bound on $\Sigma$ using all linearly independent microscopic vorticity currents. By implementing a state-space contraction, the results derived in this section hold for any choice of coarse-grained vorticity current defined between two observable state-like quantities: `effectively' non-reciprocal systems, and are closely related to the results obtained in Ref. \cite{Ohga_2023,Shiraishi_2023,Kolchinsky_2023_spectral_bounds,Liang_2023}.
\section{Conclusion and Outlook}\label{sec:conclusion_outlook}
We have presented a unified framework of the minimum action principle (MinAP) for the entropy production rate (EPR) of discrete-state systems. By deriving an exact stochastic path integral representation of discrete-state transition dynamics, which is equal to exponentiated action and incorporates non-Gaussian transition fluctuations/effective drivings, which results in an exact non-quadratic dissipation function. This formulation provides a physical interpretation of the action Lagrangian as mean inferred EPR, analogous to the role of the energy functional in the equilibrium Boltzmann distribution. This generalization allows us to formulate a far-from-equilibrium analog of the canonical ensemble that relates EPR to transition-space mean currents and its variances and defines the thermodynamic length (TL) of microscopic currents, which are linked through the exact non-quadratic dissipation function. Using this, we derive an exact non-quadratic large deviation rate functional, which tightens the bounds on EP/EPR compared to previous close-to-equilibrium Gaussian (quadratic) and far-from-equilibrium Hessian formulations, which physically correspond to quadratic Thermodynamic-kinetic uncertainty relation and non-equilibrium linear-response. We show that the variational formulation derived here is equivalent to the Information geometric formulation, extending the applicability of Information geometric methodologies to Stochastic thermodynamics, provided thermodynamic consistency is ensured.

Using TL, we show that the non-quadratic thermodynamic-kinetic uncertainty relation (TKUR) and the fluctuation relation (FR) are manifestations of the MinAP as thermodynamic inference and partial control descriptions, respectively. This unifies FR and non-quadratic TKUR within a single framework. Moreover, we extend the applicability of MinAP to coarse-grained observable currents, making it applicable to practically accessible experimental setups/systems. The variational formulation is also particularly helpful for implementing numerical optimization in cases where an analytical solution cannot be obtained. Although our framework is developed for discrete-state systems modeled by graphs \cite{schnakenberg_1976}, it is easily extended to hypergraphs that model other physical systems, for example, nonlinear chemical reaction networks \cite{kobayashi_2023_information_graphs_hypergraphs}. This work lays the foundation for practical applications of the minimum action principle in stochastic thermodynamics of far-from-equilibrium systems. For example, the generalized finite-time optimal control framework for discrete-state systems is developed in Ref.\cite{atm_2025_gftoc}. 

\subsection*{ACKNOWLEDGMENTS}

The authors gratefully acknowledge an anonymous referee whose insightful comments and constructive suggestions improved this work and broadened its applicability to fEQ systems beyond stochastic thermodynamics.
\subsection*{References}
\bibliography{reference}
\vskip2cm
\end{document}

%% file: abbrev.tex
\newcommand{\crtlong}[1]{\hat{\eta}^\dag_{#1}} 
\newcommand{\anhlong}[1]{\hat{\eta}_{#1}}